\documentclass[epj]{SVJour}
\usepackage{amssymb,epsf}

\newcommand{\bu}{{\bf u}}

\newcommand{\br}{{\bf r}}
\newcommand{\bR}{{\bf R}}

\newcommand{\ds}{\displaystyle}
\newcommand{\cH}{{\cal H}}

\newcommand{\placefigure}[3]
{
 \begin{figure}[bt]
 \begin{center}
  \leavevmode
  \epsfxsize=#2
  \epsfbox{#1.ps}
 \end{center}
  \caption{\label{#1} #3}
 \end{figure}
}

\begin{document}

\title{
Van der Waals interaction between flux lines in High-$T_c$ Superconductors} 

\dedication{Dedicated to Johannes Zittartz on the occasion of 
  his 60th birthday.}

\author{Andreas Volmer 
  \thanks{Email address: {\tt av@thp.uni-koeln.de}}
  \inst{1} \and 
  Sutapa Mukherji \inst{2} \and
  Thomas Nattermann \inst{1}\inst{3}
}

\institute{
  Institut f\"ur theoretische Physik, Universit\"at zu K\"oln,
  Z\"ulpicher Str.\ 77, D-50937 K\"oln  
    \and
  Fachbereich Physik, Universit\"at Gesamthochschule Essen, D-45117
  Essen
    \and
  Laboratoire de Physique th\'{e}orique,
  Ecole Normale Sup\'{e}rieure, 
  24, rue Lhomond,   F-75231 Paris Cedex 05  
}

\date{Subm. on February 9, 1998}

\abstract{ In anisotropic or layered superconductors thermal
  fluctuations as well as impurities induce a van der Waals (vdW)
  attraction between flux lines, as has recently been shown by Blatter
  and Geshkenbein in the thermal case [Phys.\ Rev.\ Lett.\ 77, 4958
  (1996)] and by Mukherji and Nattermann in the disorder dominated
  case [Phys.\ Rev.\ Lett.\ 79, 139 (1997)].  This attraction together
  with the entropic or disorder induced repulsion has interesting
  consequences for the low field phase diagram. We present two
  derivations of the vdW attraction, one of which is based on an
  intuitive picture, the other one following from a systematic
  expansion of the free energy of two interacting flux lines. Both the
  thermal and the disorder dominated case are considered. In the
  thermal case in the absence of disorder, we use scaling arguments as
  well as a functional renormalization of the vortex-vortex
  interaction energy to calculate the effective Gibbs free energy on
  the scale of the mean flux line distance. We discuss the resulting
  low field phase diagram and make quantitative predictions for pure
  BiSCCO (Bi$_2$Sr$_2$CaCu$_2$O$_{8}$). In the case with impurities,
  the Gibbs free energy is calculated on the basis of scaling
  arguments, allowing for a semi-quantitative discussion of the
  low-field, low-temperature phase diagram in the presence of
  impurities.  
}

\maketitle

\section{Introduction}

Conventional type--II superconductors show in addition to the flux
repulsing Meissner state a second superconducting (Abrikosov) phase in
which the magnetic induction ${\bf B}$ enters the material in the form
of quantized flux lines (FLs) which form a triangular lattice.  Each
FL carries a unit flux quantum $\Phi_0= hc/2e$. The Abrikosov lattice
is characterized by a non--zero shear modulus $c_{66}$, which vanishes
at the upper and lower critical field, $H_{c_2}$ and $H_{c_1}$, where
continuous transitions to the normal and the Meissner state,
respectively, occur.  In his mean--field solution Abrikosov treats FLs
as stiff rods.  Close to the lower critical field $H_{c_1}$, their
interaction becomes exponentially weak and hence the FL density
$\ell^{-2}=B/\Phi_0$ vanishes as $|\ln\tilde{h}|^{-2}$ where
$\tilde{h}=(H-H_{c_1})/H_{c_1}$ denotes the reduced field strength
\cite{tink}.

Thermal fluctuations roughen the FLs resulting in a possible melting
of the Abrikosov lattice close to $H_{c_1}$ and $H_{c_2}$,
respectively, because of the softening of $c_{66}$.  This applies in
particular to high--$T_c$ materials with their elevated transition
temperatures and their pronounced layer structures \cite{blatrev}.  At
present, it is not clear whether the transition to the normal phase at
high field happens in these materials via one or two transitions.
However, melting of the FL lattice has clearly been observed
experimentally \cite{gammel}.

At low fields a first order melting transition to a liquid phase and a
change in the critical behavior of $B$ has been predicted some time
ago by Nelson\cite{Nelson}.  Quantitatively the influence of thermal
fluctuations is described by a thermal length scale
$L_T=\Phi_0^2/(16\pi^2T)\approx 2\,{\rm cm}\,{\rm K}/T$ \cite{fish}.
$L_T$ has a simple physical meaning: an isolated flux line segment of
length $L_T$ shows a thermal mean square displacement of the order of
the London penetration length $\lambda$.  Besides a shift of
$H_{c_1}$, large scale thermal fluctuations lead close to $H_{c_1}$ to
an entropic repulsion $\sim (\lambda^2/L_T\ell)^2$ between FLs which
dominates over the bare interaction for small $\tilde{h}$ and hence $B
\sim \tilde{h}$ \cite{Nelson}. Here and below, all FL interactions are
measured in units of $\varepsilon_\circ = (\Phi_0/4\pi\lambda )^2=
L_{T} T/{\lambda}^2$.

More recently, Blatter and Geshkenbein \cite{BlatterGeshkenbein} found
that in anisotropic or layered superconductors short scale
fluctuations give rise also to an {\it attractive} van der Waals (vdW)
interaction \cite{brandt}.  For FLs separated by a distance $R$ the
strength of this interaction is of the order $-\lambda^6/(dL_T R^4)$
for $\lambda<R<d/\varepsilon$ and of the order
$-\lambda^6/(\varepsilon L_T R^5)$ for
$d/\varepsilon<R<\lambda/\varepsilon$.  $\varepsilon^2 =m/M \ll 1$
denotes the anisotropy of the material with $m$ and $M$ the effective
masses parallel and perpendicular to the $ab$ plane, and $d$ the
interlayer spacing.  $\lambda$ and $\lambda /\varepsilon$ are then the
screening lengths parallel and perpendicular to the layers,
respectively.

The competition among  the bare, the entropic and the vdW interactions 
leads to an interesting phase diagram at low $B$ values.
In particular, Blatter and Geshkenbein \cite{BlatterGeshkenbein}
 find at low $T$ a first order transition
between the Meissner and the Abrikosov phase.

So far fluctuation effects have been discussed for a clean
superconductor.  It is well known, however, that in type--II
superconductors FLs must be pinned in order to prevent dissipation
from their motion under the influence of an external current.
Therefore, besides the thermal fluctuations one has to take into
account the effect of disorder.  Randomly distributed pinning centers
lead indeed to a destruction of the Abrikosov lattice \cite{larkin},
but as has been recently shown, for not too strong disorder FLs may
form a (Bragg--) glass phase which is characterized by quasi
long--range order of the FL lattice \cite{bragg}.  For low $B$ this
phase undergoes a melting transition to a pinned liquid state.  Inside
this phase, disorder induced effects are expected to dominate over
those of thermal fluctuations for sufficiently low $T$.  The influence
of the disorder fluctuation induced forces between the FLs and the
consequences for the low $B$ phase diagram -- which deviates
substantially from that found in Ref. \cite{BlatterGeshkenbein} for
pure systems -- have been considered in \cite{NattermannMukherji}.

In this paper, we address several issues.  We start with a review of
the interactions between FLs and the derivation of the properties of
an isolated FL in section \ref{ChainSection}. Correlation functions,
both in the thermal and in the disordered case, are calculated. These
are needed for the calculation of the van der Waals interaction in
section \ref{vdW_sec}. There, we first develop an intuitive picture to
derive the vdW energy driven by thermal fluctuations and by
disorder. In the presence of strong disorder, detailed level
statistics of the random impurity distribution are necessary. In the
second part of this section, a thorough derivation of the van der
Waals interaction by a systematic expansion of the free energy of two
fluctuating FLs \cite{BlatterGeshkenbein,NattermannMukherji} is
presented.

The consequences for the low field phase diagram of
anisotropic high-$T_c$ superconductors are considered in section
\ref{phasediag_sec}.  In the thermal case, a functional
renormalization group is used to calculate the effective Gibbs free
energy on the scale of the mean distance $\ell$ between FLs. In particular,
the bare interaction between flux lines, which is given by the
superposition of the bare repulsion and the vdW attraction, is
renormalized by integrating out thermal fluctuations on scales between
$\lambda$ and $\ell$.  The results of this calculation are compared to
an expression for the Gibbs free energy based on scaling arguments. In
this latter approach, the contribution from the vdW interaction can be
estimated only up a to numerical factor determining its amplitude;
this factor will be quantified using the results from the
renormalization procedure.  In the case with disorder, we give an
expression for the Gibbs free energy that is based on scaling
arguments, and which allows for a semi-quantitative discussion for the
low-field, low-temperature phase diagram in the presence of
impurities.

\section{Single chain properties}
\label{ChainSection}

In this section we discuss properties of an isolated flux line (FL).
Starting with a vortex lattice picture with the general form of the
interaction between the FLs, we consider the limit of a very dilute
lattice where a single chain approximation is good enough. A single FL
in the dilute limit is well described by a dispersive stiffness
constant $\varepsilon_l(k_z)$, the origin and different limiting
properties of which are discussed. Deriving the Hamiltonian for a
FL in an isotropic material, we generalize it for a FL in the presence
of impurities with its full dispersive stiffness constant. Finally, we
discuss the properties of several correlation functions in the
presence of impurities. These correlation functions will be used in 
the derivation of the van der Waals interaction in section \ref{vdW_sec}.

\subsection{Interactions and elastic constants}
\label{sec:ElasticConstants}

In reality the flux lines  are not  straight; they are distorted from
their  equilibrium positions. The displacement with  respect to the
equilibrium position ${\bf r}_i\equiv ({\bf a}_i,z)$, where $i$
denotes a site on the planar lattice, is  represented by a two
dimensional vector ${\bf u}_i(z)$ \cite{blatrev}. 
In the low field limit, where the London theory is applicable, the
interaction energy between $N$ vortices  is in general  given by 
\begin{equation}
  {\cal F}=\frac{\varepsilon_\circ}{2}\sum_{i,j=1}^N
  \int d{ s}_{i\alpha}\  d{s}_{j\beta}\ 
  V^{\rm int}_{\alpha\beta}( {\bf s}_i-{\bf s}_j)
\label{iso},
\end{equation}
where ${\bf s}_i={\bf r}_i+{\bf u}_i(z)$ is the position vector of the
distorted vortex segment. $(\alpha \beta)\in (x,y,z)$ indicate 
different components. 
In (\ref{iso}) the terms with $i=j$ have been included. These terms
correspond to the interaction between the segments of the same vortex
line and therefore contribute to the self energy of the line. 
In isotropic materials the interaction is exponentially
damped and is of the form ${e^{-r/\lambda}}/{r}$ with a short
distance cutoff at the coherence length $r=\xi$.
The high temperature oxide superconductors are essentially layered
with the conducting CuO planes parallel  to the $ab$ plane. A large
anisotropy between the $c$ 
axis and the $ab$ plane due to this layered structure leads to many
unusual properties compared to the isotropic materials.

The tensorial London potential between the vortices is conventionally
expressed through its Fourier transform 
defined as 
$ V^{\rm int}_{\alpha\beta}({\bf r})=4\pi\lambda^2 \int d^3k/(2\pi)^3\,
e^{i{\bf kr}} {\tilde V}^{\rm int}_{\alpha\beta}({\bf k})$
\cite{brandt3}. In our case, however, it turns out to be more
convenient to express it in terms of its partial Fourier transform
with respect to the $z$-direction only, as defined by
\begin{equation}
  V_{\alpha\beta}^{\rm int}(\bR,k_z) = 
  \frac{1}{4\pi\lambda^2} \int dz \; e^{-ik_zz}\,
  V_{\alpha\beta}(\bR,z),
\end{equation}
where ${\bf r}=(\bR,z)$.
This choice has the advantage of an explicit dependence on the distance
  $\bR$ between FLs. 
Choosing $\bR=(R,0)$ and $R\gtrsim \lambda$,
the only contributions to the tensorial London potential 
that do not vanish by symmetry are
\begin{equation}
\label{V_zz}
  V_{zz}^{\rm int}(\bR,k_z) = -\frac{1}{2\pi\lambda^2}\;
  K_0\left(\sqrt{1+\lambda^2 k_z^2}\,R/\lambda\right)
\end{equation}
and
\begin{eqnarray}
\label{V_xx}
  &&V_{xx}^{\rm int}(\bR,k_z) = -V_{yy}^{\rm int}(\bR,k_z) \\
  &&\qquad= -\frac{1}{2\pi R} \frac{\varepsilon}{\lambda\sqrt{1+\lambda^2
  k_z^2}} \, K_1\left(\varepsilon\sqrt{1+\lambda^2
  k_z^2}\,R/\lambda\right).
\nonumber
\end{eqnarray} 
$K_n(x)$ is the modified Bessel function of the second kind of
$n^{\rm th}$ order. 
In the extreme anisotropy limit $\varepsilon\rightarrow 0$, this
latter equation (\ref{V_xx}) simplifies to
\begin{equation}
\label{V_xx_eps0}
  V_{xx}^{\rm int}(\bR,k_z) = -\frac{1}{2\pi\,R^2}\,
  \frac{1}{1+\lambda^2 k_z^2}.
\end{equation}
In (\ref{V_zz}) to (\ref{V_xx_eps0}), a short distance cutoff at the
coherence length $\xi$ is implied. 
In (\ref{V_xx}) and (\ref{V_xx_eps0}) subdominant terms of
order $e^{-R/\lambda}$ have been neglected.

Different elastic moduli of the vortex lattice 
can be obtained from (\ref{iso}) by
expanding in small displacement ${\bf u}$. For very small field 
$H\lesssim H_{c_1}$, the shear and compression moduli $c_{66}$
and $c_{11}(k)$
decay exponentially with the lattice
spacing $\ell$ and can be neglected in a very dilute limit. 
The tilt modulus $c_{44}(k)$ is only weakly dispersive in isotropic
materials \cite{brandt4}. In anisotropic materials, though, it has a much
stronger dispersivity \cite{brandt1,blatrev} which results, in the single FL
limit, in a strongly dispersive elastic stiffness
constant $\varepsilon_l(k_z)=\ell^2 c^c_{44}(k_z)$
consisting of two parts,
\begin{equation}
\varepsilon_l(k_z)\approx
\frac{\varepsilon^2 \varepsilon_\circ}{2}
\ln\left(\frac{\kappa_c^2}{1+\lambda^2
    k_z^2}\right)+\frac{\varepsilon_\circ}{2\lambda^2
    k_z^2}\ln(1+k_z^2\lambda^2),
\label{stiff}
\end{equation}
where $\kappa_c=\kappa/\varepsilon$ with the Ginzburg-Landau parameter
$\kappa=\lambda/\xi$. $c^c_{44}(k_z)$ is the part of the tilt
modulus that originates from the self-energy of the vortices.
The first part in (\ref{stiff}) represents the Josephson coupling
between the layers and is absent in the limit $\varepsilon\rightarrow
0$. The second part describes the contribution from
the electromagnetic coupling, which remains finite in this limit.  In
the extreme short wavelength regime $\lambda k_z>1/\varepsilon$ 
the contribution due to the  Josephson coupling 
 dominates over the electromagnetic part and the line stiffness is
given by the first term in (\ref{stiff}).
In the intermediate regime 
$1/\lambda<k_z<1/\varepsilon\lambda$, the electromagnetic interaction
dominates, and the elastic constant is given by the second term in
(\ref{stiff}), hence the stiffness is highly dispersive in this regime.
In the long wavelength limit, the line stiffness reaches a
constant, 
\begin{equation}
\label{stiffLongWave}
  \varepsilon_l(k_z)\approx \frac{\varepsilon_\circ}{2}\,
  (1+\varepsilon^2 \ln\kappa_c^2) \qquad\mbox{for }\lambda k_z\ll 1.  
\end{equation}

In order to obtain an estimate for the contributions from the
Josephson and the electromagnetic coupling to the stiffness constant
at small length scales we use $k_z=\pi/d$, where the layer spacing $d$
is the lowest relevant length scale. Using the parameters
$\lambda=2000\,${\AA}, $\xi=20\,${\AA}, $d=15\,${\AA} and
$\varepsilon=1/300$ \cite{blatrev}, suitable for BiSCCO
(Bi$_2$\-Sr$_2$\-Ca\-Cu$_2$\-O$_{8}$), we find that the contributions
from the electromagnetic and the Josephson couplings are about equal
in magnitude at this wavelength, while for lower $k_z$, the electromagnetic
part dominates. Hence, for BiSCCO the Josephson coupling may be
entirely neglected. For materials with a larger value of
$\varepsilon$, on the other hand, $k_z=\pi/d$ belongs to the extreme
short wavelength limit ($\lambda k_z>1/\varepsilon$) where the weakly
dispersive Josephson coupling dominates.

\subsection{Hamiltonian for a single Flux line}
\label{sec:SingleFLHamiltonian}

The line stiffness $\varepsilon_l(k_z)$
obtained above can be used to describe FLs as
elastic strings with a strongly dispersive elastic constant.  In
practice it is difficult to predict the properties of the FL 
at all wavelengths. Often in the following we
therefore aim at large or small wavelength features depending on our
interest in distortions on scales larger or smaller compared to
$\lambda$.  In 
the extreme long wavelength case where FLs are elastic strings with a
dispersionless stiffness constant, the underlying lattice structure can
be ignored. In the opposite limit the layered structure and the
anisotropy become important \cite{blatrev}. Depending on the ratio
$\tau_c=2 \xi_c^2/d^2$, where $\xi_c=\varepsilon \xi$ 
is the coherence length in the direction parallel to
the $c$ axis with $\varepsilon$ denoting the anisotropy parameter,
one can distinguish two cases, the continuous
anisotropic limit and the extreme decoupled discrete limit. For
$\tau_c\gg 1$ the coherence length is large enough to neglect the
discrete layered structure. In that case the continuous anisotropic
Ginzburg-Landau description is applicable.  $\tau_c\ll 1$ is the
decoupled limit where the Lawrence-Doniach model \cite{lawrence} is more
appropriate.  In the latter case, instead of the continuous description,
the vortex is often viewed as a stack of pancake vortices residing in
layers but interconnected by Josephson strings between two successive
CuO planes, as sketched in Fig.~\ref{pancakes}.  The effect of the
thermal energy or of random impurities 
is to displace the pancakes from their aligned position. Due to
the dispersion these local displacements and their correlations
will be different from the long wavelength distortions.  It is however
important to point out that the continuous anisotropic Ginzburg-Landau
theory  often provides a good description of the layered structure
\cite{brandt2} unless specific thermodynamic properties, where the 2d
structure is more important than the 3d bulk material, are
investigated.

\placefigure{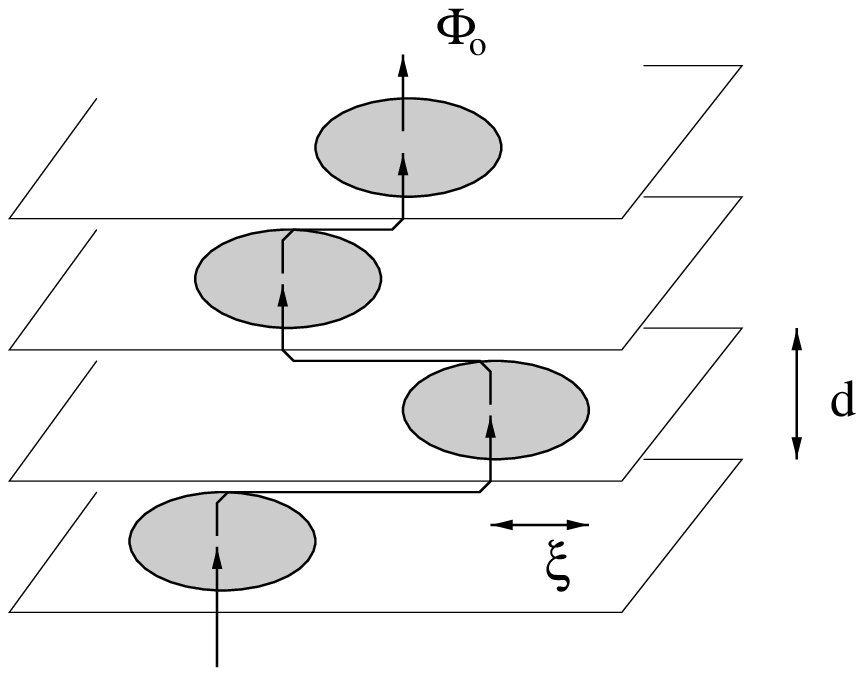}{0.4\textwidth}{A single vortex line, carrying
  the flux $\Phi_0$, in a strongly layered superconductor. The flux
  line consists of an array of pancake vortices threading the
  individual superconducting layers which are separated by a distance
  $d$; the pancakes are interconnected by Josephson strings.}

Taking into account the dispersivity of the stiffness constant of
the FL, the Hamiltonian for a single FL in
the presence of random impurities can be defined as  
\begin{eqnarray}
  {\cal H}(\{\bu\})=
  \sum_k\ \frac{\varepsilon_l(k)}{2}\, k^2 \,
  \tilde{\bu}_k \tilde{\bu}_{-k}+ 
  \int_0^L  dz\, \varepsilon_{\rm pin}({\bf u}(z),z).
\label{mutual}
\end{eqnarray}
Here only the bending energy part has been expressed in momentum
space with the Fourier transform of ${\bf u}(z)$ defined as 
\begin{equation}
  {\tilde{\bf u}}_k=\frac{1}{\sqrt {L}}\int_0^L dz\ e^{ikz}{\bf u}(z).   
\label{ft}
\end{equation}
Here and below, $k$ denotes the $k_z$-component of the full wave
vector ${\bf k}$.
 
The random pinning potential is assumed to be Gaussian distributed
with $\overline{\varepsilon_{\rm pin}(\bf u)}=0$ 
(a nonzero mean value $\overline{\varepsilon_{\rm pin}}$ simply shifts
the 
Hamiltonian by a constant and is hence inconsequential)
and 
\begin{eqnarray}
  \overline{
    \varepsilon_{\rm pin}({\bf u},z)\varepsilon_{\rm pin}({\bf 0},0)
    } \;=\;
  \frac{T^3_{\rm dis}}{\varepsilon_\circ{\xi^2}}\;
  \delta(z)\tilde{k}({\bf u}/\xi) 
  \label{correl} 
\end{eqnarray}
with $\tilde{k}(x)=1$ for $x\ll 1$ and $\tilde{k}(x)\approx (1/x^2)\ln
x$ for $x\gg 1$, respectively \cite{blatrev}.  The overbar denotes the
average over the disorder distribution. The displacement caused by the
quenched disorder is strongly suppressed due to thermal smoothening
for $T>T_{\rm dis}$.  The characteristic energy $T_{\rm dis}$ can be
related to the local shift in the critical temperature $T_c$ which is
induced by an oxygen impurity density $n_i$ by
\begin{equation}
  \frac{T_{\rm dis}^3}{(\varepsilon_\circ\xi)^3} =
  0.4\left( \frac{n_i}{T_c} \frac{dT_c}{dn_i} \right)^2
  \frac1{\xi^3n_i} \frac1{(1-t)^{1/2}},
\end{equation}
where $t=T/T_c$ \cite{fish,blatrev}. This result holds
in the case of non-optimal doping where $dT_c/dn_i\neq 0$; if this
derivative vanishes, $T_{\rm dis}$ can be similarly related to 
the second derivative $d^2T_c / dn_i^2$ \cite{blatrev}.

Statistical mechanics of a single FL in a random potential shows that
the disorder is always relevant for $d\leq 2$. This is reflected
through an enhanced averaged  FL wandering on large scales 
$L>L_{\rm dis}$ 
\begin{equation}
  \overline{\langle \,[\,{\bf u}(L)-{\bf u}(0)\,]^2\,\rangle}
  \sim \lambda (L/L_{\rm dis})^\zeta 
\label{displace}
\end{equation}
with $\zeta>1/2$ for $d\leq 5$ \cite{kinzel}. Here and in the
following the angular bracket indicates thermal average. Several exact
treatments at $d=2$ lead to $\zeta=2/3$. Disorder is marginally
relevant at $d=3$. Numerical simulations  and recent analytical
calculations show  $\zeta=5/8$ \cite{tang,lassig}. In the presence of
disorder vortex segments of length $L_{\rm dis}$ are independently
pinned. At $T=0$, this can be seen by minimizing the energy
consisting  of elastic and pinning contributions. Here the pinning
energy is estimated within perturbation theory. This provides 
\begin{equation}
\label{lowtemp}
  L_{\rm dis}=\varepsilon_\circ \lambda^2 
  \kappa^{1/\zeta-2}/ T_{\rm dis}. 
\end{equation}
For $T<T_{\rm dis}$,  the thermal fluctuations can be ignored 
and the above estimate of $L_{\rm dis}$ is still
valid. For $T>T_{\rm dis}$, on the other hand,
\begin{equation}
\label{hightemp}
  L_{\rm dis}(T) =
  \frac{\varepsilon_\circ \xi^2}{T}
  \exp\left[\left(T/T_{\rm dis}\right)^3\right]
\end{equation} 
shows
an exponential growth with the temperature as a consequence of the
marginal relevance of the disorder at 3 space dimension
\cite{NelsonDoussal}. 
In the pure thermal case $\zeta=1/2$, and $L_{\rm dis}$ is replaced by
 \cite{fish}
\begin{equation}
\label{L_T}
  L_T= \varepsilon_\circ \lambda^2/T \; .
\end{equation}

\subsection{Correlation functions}
\label{sec:CorrFunc}

With the ideas sketched above we now study several correlation
functions in the presence of quenched impurities. Whereas in a pure
system $\langle {\bf u}_k\rangle=0$ and hence $\langle {\bf u}_k {\bf
  u}_{-k}\rangle$ is the appropriate correlation function, in a random
system one has to  distinguish between the connected correlation
function 
\begin{equation}
\label{C_T}
  C_T(k)={\overline {\langle {\bf u}_{k} {\bf u}_{-k}\rangle}}
  -{\overline{ \langle {\bf u}_{k}\rangle \langle
{\bf u}_{-k}\rangle}}
\end{equation}
and the disconnected  correlation function 
\begin{equation}
\label{C_dis}
  C_{\rm dis}(k)={\overline
    {\langle {\bf u}_k\rangle \langle {\bf u}_{-k}\rangle}}.
\end{equation}
We first study $C_{\rm dis}(k)$ and near the end of this
section we show that the disorder does not affect $C_T(k)$ obtained in
the pure case.  

\subsubsection{Short wavelength limit of $C_{\rm dis}(k)$}

We first consider the short wavelength 
behavior of $C_{\rm dis}(k)$. In this limit the problem essentially 
simplifies to obtaining the typical displacement of a single pancake.
A perturbative technique and an Imry-Ma type argument are applied
to estimate the single pancake displacement.

As mentioned above in section \ref{sec:ElasticConstants}, for highly
anisotropic superconductors with very small values of $\varepsilon\ll
1$ (such as BiSCCO), the highest relevant wave vector $\pi/d$ is
smaller than $1/\lambda\varepsilon$, with the consequence that in the
short wavelength limit $\lambda k\gg 1$ it is sufficient to consider
the dispersive elastic constant
$\varepsilon_l(k)=(\varepsilon_\circ/\lambda^2 k^2) \ln(\lambda
k)$. This term represents the contribution from the electromagnetic
coupling in the full expression (\ref{stiff}) for the line stiffness.
A closer look at the elastic energy term in (\ref{mutual}) with the
dispersive elastic constant reveals
that, apart from a logarithmic factor,
the pancakes on different layers are essentially uncoupled. 
Thus rewriting the correlation function in real space 
\begin{equation}
{\overline{\langle {\bf u}_k\rangle \langle {\bf u}_{-k}\rangle}}
=\frac{1}{L}\int dz\ dz' e^{-ik(z-z')} \overline{\langle {\bf u}(z)\rangle
    \langle {\bf u}(z')\rangle}, 
\end{equation}
we utilize the above assumption of independent pancake displacement,
i.e. 
\begin{equation}
\label{correl1}
  \overline{\langle {\bf u}(z)\rangle \langle {\bf
      u}(z')\rangle}=u_{\rm pv}^2 f(z-z'),
\end{equation}
where
\begin{eqnarray}
  f(z)= \left\{
    \begin{array}{cl}
      1 & \mbox{ for } |z|<d/2 \nonumber \\
      \rule{0cm}{3ex}
      0 & \mbox{ for } d/2<|z|\ll\lambda.
    \end{array}
  \right.
\end{eqnarray}
This leads to 
\begin{equation}
  \label{pvdis}
  C_{\rm dis}(k)\approx {\overline{\langle u_{\rm pv}^2 \rangle}}\,d
  \qquad \mbox{for}\quad 1\ll \lambda k\lesssim \pi\lambda/d.
\end{equation}
The mean square
displacement of a pancake in the presence of quenched impurities can be
obtained from the Hamiltonian  (\ref{mutual}) 
which is, in the short wavelength limit, a superposition of the
following two parts: (i) a parabolic elastic potential,
$E_{\rm el}(u)\approx (u^2d/2)\,k^2\varepsilon_l(k)|_{k=\pi/d}$,
which in the case considered here (where the electromagnetic coupling
dominates) reads
\begin{equation}
  E_{\rm el}(u)=\frac12 
  \frac{\varepsilon_\circ d}{\kappa^2} \ln(\pi\lambda/d)
  \frac{u^2}{\xi^2} ,
\label{E_el}
\end{equation}
and (ii) a contribution from the disorder, averaged over the pancake
height $d$, 
\begin{equation}
\label{epspin_d}
  \varepsilon_{\rm pin}(\bu) \equiv
  \int_0^d dz\ \varepsilon_{\rm
  pin}({\bf r}+\bu(z),z),
\end{equation}
where ${\bf r}$ is the equilibrium position of the pancake.

It is useful to re-express this Hamiltonian 
in dimensionless variables $t=z/d$ and
$\hat\bu=\bu/\xi$. In terms of these dimensionless variables,
\begin{eqnarray}
\label{Hdimless}
  \frac{{\cal H}}{T}&\approx &
  \frac{\varepsilon_\circ d}{T\kappa^2}
 \ln(\pi \lambda/d)\, \times \\
 && 
 \left\{    
   \frac{\hat\bu^2}{2} +
   \Delta^{1/2}(\pi\lambda/d)
   \int_0^1 \!\!\!dt\ \tilde
   \varepsilon_{\rm pin}(\hat\bu,t)
  \right\}, \nonumber
\end{eqnarray}
where ${\tilde\varepsilon}_{\rm pin}(\hat\bu,t)$ is the pinning
potential scaled to have a variance equal to unity, and we have defined
\begin{equation}
\label{delta}
  \Delta(x)=\Delta_0\,  \frac{x}{\ln^2(1+x^2)},
  \qquad
  \Delta_0 =   
  \left( \frac{T_{\rm dis}\kappa^2}{\varepsilon_\circ\lambda}
  \right)^3.
\end{equation}
From (\ref{Hdimless}) one concludes that in general, at finite
temperature and finite disorder strength,
\begin{equation}
  \label{gendis}
  {\overline{  \langle u^2_{\rm pv} \rangle}}=\xi^2
    \Phi\left(\Delta(\pi\lambda/d),\;\frac{T 
      d}{\xi^2\varepsilon_l(\pi/d) }\right),
\end{equation}
where
and $\Phi$ is a function of dimensionless variables for
the disorder and the temperature. 

In the absence of disorder ($\Delta_0=0$) fluctuations are driven by 
thermal energy and we have
\begin{equation}
\label{therm}
  \frac{\langle u^2_{\rm pv}\rangle_{\rm th}}{\xi^2} = 
  \frac{2T}{\varepsilon_\circ d }\,
  \frac{\kappa^2}{\ln (\pi \lambda/d)}.
\end{equation}
At $T=0$ and finite disorder $(T_{\rm dis}>0)$, the FL fluctuates
to take advantage of the impurities, and the ground state displacement
in this case is 
\begin{equation}
  {\overline {u_{\rm pv}^2}}/{\xi^2}=F(\Delta(\pi\lambda/d))
  \equiv \Phi(\Delta(\pi \lambda/d),0).
\end{equation}
For very low temperatures $T\ll T_{\rm dis}$,
we can approximate the mean square displacement as 
\begin{equation}
  \label{impudis}
  \overline{\langle\bu_{\rm pv}^2\rangle}
  \approx \overline{\bu_{\rm pv}^2}= 
  \xi^2 F(\Delta(\pi\lambda/d)).
\end{equation}
In order to determine the explicit functional form of $F(\Delta)$, we
take a recourse to a perturbative approach valid for small
displacements $u\ll\xi$. In order to go beyond the perturbative regime,
an Imry-Ma type scaling argument and the result from
numerical simulations are discussed. 
In general we expect a power law form of 
\begin{equation} 
\label{eta}
  F(\Delta)\sim \Delta^\eta
\end{equation}
and we attempt to get the value of $\eta$.

In the perturbative scheme we proceed with the reduced
Hamiltonian (\ref{Hdimless}) for a single pancake vortex.
Expansion of the pinning energy in small
displacement $\hat\bu$
\begin{equation}
  \tilde\varepsilon_{\rm pin}(\hat\bu,t)=
  \tilde\varepsilon_{\rm pin}(0,t)+
  \hat\bu\cdot\nabla\tilde\varepsilon_{\rm pin}
    (\hat\bu,t)\big|_{\hat\bu=0}+\ldots 
\end{equation}
and minimization of the total energy with respect to the
displacement leads to the force equation 
\begin{equation}
  \hat\bu =
  -\Delta^{1/2}(\pi\lambda/d)
  \int_0^1 dt\, \nabla\tilde\varepsilon_{\rm pin}
    (\hat\bu,t)\big|_{\hat\bu=0} \; .
\label{forcebalance_pert}
\end{equation}
Averaging over the dimensionless distribution 
$\tilde\varepsilon_{\rm pin}(\hat\bu,t)$ directly leads to
the disorder induced short length scale displacement
\begin{equation}
  \overline{u_{\rm pv}^2} \simeq \xi^2\Delta(\pi\lambda/d)
 \label{pert5}.
\end{equation}
Thus, within the perturbative approach $\eta=1$.

Due to the  narrow range of applicability of the perturbation  
technique, a more general nonperturbative scheme is needed 
for $\Delta(\pi\lambda/d)\gg 1$. This can be done by an Imry-Ma or
variational approach \cite{engel}, which 
briefly proceeds as follows. Consider the dimensionless part (in
curly brackets) of the single pancake Hamiltonian
(\ref{Hdimless}). The displacement of a single pancake
by an amount $\hat\bu$ is associated
with the elastic energy cost $\hat\bu^2/2$. This energy cost has to be
compared with the possible gain in
pinning energy corresponding to the adjustment of the pancake position
to a minimum of the random potential within the area $\pi\hat\bu^2$.
This energy is obtained from the second term in (\ref{Hdimless}) and
is of order $\Delta^{1/2}(\pi\lambda/d)$, where we have used that
$\tilde\varepsilon(\hat\bu,t)$ has been scaled to unity, i.e.
\begin{equation}
  \overline{
    \int_0^1 dt\,\tilde\varepsilon_{\rm pin}(\hat\bu,t)
    \int_0^1 dt'\,\tilde\varepsilon_{\rm pin}(\hat\bu',t')}
  \;=\; 1.
  \label{disorderscale}
\end{equation}
Balancing the two energies we find 
\begin{equation}
  \label{ImryMa}
  \overline{ u_{\rm pv}^2 }
  \sim \xi^2 \, \Delta^{1/2}(\pi\lambda/d),  
\end{equation}
corresponding to the exponent $\eta=1/2$. 

In the derivation above we have neglected the fact that the gain in
pinning energy is not independent of $\hat\bu$. With
$\tilde\varepsilon(\hat\bu,t)$ obeying a Gaussian distribution, it is
in fact proportional to $\ln^{1/2}(\hat\bu^2/2\sqrt\pi)$
\cite{koshlev}. Taking account of this dependency leads to a confluent
logarithmic term of the form $F(\Delta)\sim
\Delta^{1/2}\ln^{-1/2}(\Delta)$.  Nevertheless, we will only use the
leading scaling behavior of $F(\Delta)$ in the rest of the paper.

A numerical estimate for BiSCCO with the parameters as used before and
$T_{\rm dis}=45\,$K yields $\Delta(\pi\lambda/d)\approx 300$, hence
the Imry-Ma result should apply, predicting a typical displacement
$\overline{u_{\rm pv}^2}/\xi^2$ of order 20.

Finally, we mention the result from
a numerical study.  We have simulated the energy on a 2-d lattice by
superposing the elastic energy quadratic in the position ${\bf r}$ on
the lattice and a random, Gaussian distributed   
energy $V(r)$ \cite{engel}. A log-log plot of the mean square position
of the minimum energy site, obtained after a large number of sample
averaging, versus the width of the distribution, shows roughly a
straight line with a slope that corresponds to an exponent
$\eta\approx 4/9$. While this may not reflect the correct {\em asymptotic}
scaling behavior, it gives a good representation of the scaling in the
regime of physically accessible values for $\Delta$, and is consistent
with the confluent logarithmic term mentioned above.

In conclusion, we find in the short wavelength and low temperature limit
\begin{equation}
  \label{cdiss}
  C_{\rm dis}(k)={\overline{u_{\rm pv}^2 }}\,d
  \qquad \mbox{for}\quad 1\ll \lambda k\lesssim \pi\lambda/d,
\end{equation}
where ${\overline{u_{\rm pv}^2}} \approx \xi^2 F(\Delta(\pi \lambda/d))$ and
$F(\Delta)\sim \Delta^\eta$ with $\eta=1$ for $\Delta\ll 1$ and $\eta\approx
1/2$ for $\Delta\gg 1$.

\subsubsection{Long wavelength limit of $C_{\rm dis}$}

In the long wavelength limit $\lambda k\ll 1$,
we are aided by the known scaling form of the mean square
displacement (\ref{displace}) and the  value of $\zeta$
 from numerical simulations at  $d=3$.
We proceed with an Imry-Ma argument as described above, with the
replacement of $\pi \lambda/d$ 
by $\lambda k$. For $\lambda k\ll 1$ we have 
\begin{equation}
  \Delta(\lambda k\ll 1)\simeq
  \Delta_0/ (\lambda k)^3.
\end{equation} 
Therefore using 
\begin{equation}
  \label{longw}
  \langle u^2 \rangle =\int _{1/L}\frac{dk}{2\pi}\ C_{\rm dis}(k)=\xi^2
  \int_{1/L}  \frac{dk}{2\pi k}\ F(\Delta(\lambda k))
\end{equation}
together with (\ref{displace}), 
we find $F(\Delta(\lambda k))\sim [\Delta(\lambda k)]^{2\zeta/3}$. 
This leads to $L_{\rm dis}$ in agreement with
(\ref{lowtemp}). The above treatment is valid for very low temperature $T\ll
T_{\rm dis}$. At higher temperatures the crossover length scale grows
exponentially as given by Eq.~(\ref{hightemp}).

\subsubsection{Thermal fluctuations $C_T$}
The other quantity of interest is the disorder averaged second
cumulant of thermal fluctuation 
$C_T(k)$ as defined in (\ref{C_dis}).
In the pure system the single vortex
fluctuation amplitude is $\langle \bu_k
\bu_{-k}\rangle=T/\varepsilon_l(k)k^2$. 
  In the presence of disorder, $C_T(k)$ remains unaltered, i.e. 
\begin{equation}
  C_T(k)=\frac{T}{\varepsilon_l(k) k^2}. 
\end{equation}
This can be readily seen from the
discretised single vortex Hamiltonian in a random potential by adding
a source term $T\sum_i \lambda_i u_i$. Derivatives of the
partition function with respect to $\lambda_i$ lead to various
correlations. The above result follows by rewriting 
the Hamiltonian with a new variable $u_k-T
\lambda_k/\varepsilon_l k^2$ in terms of which the random potential is
assumed to have the same distribution \cite{schulz}.

\section{Van der Waals attraction}
\label{vdW_sec}
In the first part of this section we discuss the origin of the van der
Waals (vdW) attraction considering a simple physical picture of
distortion of pancakes and subsequent formation of pancake-anti
pancake dipole. This simple approach correctly accounts for the vdW
attraction in the pure system. Using the same picture, we also discuss
the vdW attraction in the presence of impurities in the medium -- a
problem more complicated than the pure case due to the averaging over
the impurity distribution which has to account for the presence of
metastable states. In the second part (section \ref{sec_statmech}), we
derive the vdW attraction using the statistical mechanics of
interacting flux lines. The power law behavior of the vdW interaction
as a function of the FL distance $R$ differs for the Josephson coupled
and uncoupled cases due to the very nature of the interaction.

For a  stack of only electromagnetically coupled 
pancake vortices the interaction between two pancake vortices in the
same layer separated by a distance $R>\xi$ is given by 
$V(R)=2 d \varepsilon_\circ \ln(R/\xi)$.
Pancakes on different layers attract each other with a strength
reduced by a factor $d/\lambda$.  Josephson coupling between the layers
leads to an additional contribution to the free energy that
depends on the phase difference of the superconducting order parameter
on two successive layers.  A comparison of the Josephson and the
electromagnetic interaction energy \cite{blatrev} shows that for
separations $R<d/\varepsilon$ the latter is much stronger, whereas the
Josephson interaction dominates in the regime
$R>d/\varepsilon$.  Therefore the interaction between two Josephson
coupled pancakes in the same layer is given by
\begin{equation}
\label{joscoup}
V(R)= \left\{
  \begin{array}{ll}
    2d\varepsilon_\circ \ln\ds\frac{R}{\xi}, &
    \xi\lesssim R\lesssim d/\varepsilon\nonumber\\ 
    \rule{0cm}{5ex}
    2d\varepsilon_\circ 
    \left(
      \ds\frac{\varepsilon R}{d}-
      \frac{d}{4\varepsilon R}\right), 
    &d/\varepsilon\lesssim R\lesssim\lambda/\varepsilon . 
  \end{array}
  \right.
\end{equation}
For two straight stacks of pancakes placed at a distance
$R\gg\lambda$, the mutual attraction from different layers 
 and repulsion from the same layer  compensate each
other leaving a residual interaction $\sim \exp(-R/\lambda)$ as it is
in the Abrikosov lattice. 

\subsection{An intuitive picture}
A FL with a single pancake distorted due to thermal energy or
disorder can be visualized as a straight FL with a dipole constructed
out of the displaced pancake and an anti-pancake at the vacant position
left behind by the displaced pancake. The dipole thus formed induces
another dipole on the neighboring vortex by virtue of the interaction
between the pancakes of different vortices and a finite restoring
force of the same stack. These fluctuation induced dipoles interacting
with the long range interaction originating from (\ref{joscoup}) lead
to a long range interaction between vortices  very similar to the
quantum fluctuation induced attraction in atomic system. In analogy
with the latter system these fluctuation induced attraction are called
van der Waals interaction.  

If two pancakes in the same layer of stacks $1$ and $2$
are displaced by an amount $u_1$ and $u_2$, respectively, the
dipole-dipole interaction energy between them is given by
\begin{equation}
  U_{12}= \frac{2\varepsilon_\circ d}{R^2}[\bu_1\cdot\bu_2-
  2(\bu_1\cdot {\bf \hat n})({\bf u}_2\cdot {\bf \hat n})].
\label{U_12}
\end{equation}
Here $\hat {\bf n}$ denotes the unit vector along the line connecting
the two vortices. 
The force exerted on the pancake of stack 2 due to a dipole on stack 1
(in the same layer) is  
\begin{equation}
\label{U_12_force}
  {\bf f}_{12}=
  -\nabla_{\!\bu_2} U_{12} =
  -\frac{2\varepsilon_\circ d}{R^2}
  [\bu_1-2(\bu_1\cdot{\bf\hat n}){\bf\hat n}]. 
\end{equation}
The displaced pancake on  stack 2 on the other hand 
experiences an elastic force due to all the other pancakes in
different layers. The restoring force 
${\bf f}_{22}=-\nabla_{\bu_2}E_{\rm el}(\bu_2)$, 
where $E_{\rm el}(\bu_2)$ is the parabolic elastic potential
(\ref{E_el}) acting on the pancake vortex, is 
\begin{equation}
  {\bf f}_{22}= -\varepsilon_\circ\frac{d }{\lambda^2 }
  \ln\left(\pi\lambda/d\right){\bf u}_2.
\label{restor}
\end{equation}
In the absence of impurities, the position of pancake 2 is (in equilibrium)
exactly
determined by the force balance ${\bf f}_{12}+{\bf f}_{22}=0$, which leads 
to ${\bf u}_2=-(2\lambda^2/R^2\ln(\pi\lambda/d))
\times[\bu_1-2(\bu_1\cdot{\bf\hat n}){\bf\hat n}]$.
Substituting this result back into
(\ref{U_12}) gives the van der Waals interaction  
\begin{equation}
  V_{\rm vdw} \simeq
    -\frac{4\varepsilon_\circ}{\ln({\pi\lambda}/d)} \,
    \frac{\lambda^2}{R^4}\;
    \overline{\langle{\bu_1^2}\rangle}.
\label{vdw}
\end{equation}
Below, we will first evaluate this expression in the case of thermal
fluctuations only, and then reconsider the force balance argument in
the presence of impurities. For weak disorder, we will find that this 
argument is still valid because the random forces are much weaker than the
restoring force (\ref{restor}), while it fails in the strong disorder
regime. There, however, we will show that a different argument, using level
statistics, leads to an expression for the vdW energy that is basically
identical to (\ref{vdw}). Hence, this latter expression fully applies
both in the thermal and in the disorder dominated regime, justifying the
thermal {\em and} disorder average over $u_1^2$ which has been anticipated
here.

\subsubsection{Pure thermal case}
For the thermal case in pure systems,
we find the vdW attraction by plugging the result (\ref{therm}) for thermally
driven fluctuations into Eq.~(\ref{vdw}), which leads to
\begin{equation}
  V_{\rm vdw}^{\rm th} \simeq
  -\frac{4\varepsilon_\circ}{\ln^2(\pi\lambda/d)}
  \frac{T}{d\varepsilon_\circ} 
  \left(\frac{\lambda}{R}\right)^4 .
\end{equation}
This, apart from a numerical factor, 
 is in agreement with the vdW attraction for pure superconductors
derived in \cite{BlatterGeshkenbein}.

\subsubsection{Disorder case at $T=0$}

In the case of disorder-induced fluctuations, a more subtle treatment
is necessary.
First, let us distinguish between the weak and the strong
disorder limit, defined by $\Delta\equiv\Delta(\pi\lambda/d)\ll 1$ and
$\Delta\gg 1$, respectively. For convenience, we define the total
pinning energy $\varepsilon_{\rm pin}^{_{(i)}}(\bu_i)$ acting on pancake
vortex $i$  as in (\ref{epspin_d}).

In the weak disorder regime ($\Delta\ll 1$), the pancake position is
determined by the force equation (\ref{forcebalance_pert}), leading to
ensemble fluctuations $\overline{u_{\rm pv}^2}\simeq\xi^2\Delta$  
(\ref{pert5}) which are hence much smaller than
the disorder correlation length $\xi$. Furthermore, the restoring
forces exerted by the elastic coupling to the other pancakes in the
same vortex (\ref{restor}) are much stronger than the random forces 
$\nabla\varepsilon_{\rm pin}^{_{(i)}}(\bu_i)$. This can be seen by
estimating the ratio 
between the second derivative of the random energy and the elastic
constant $\varepsilon_\circ d\ln(\pi\lambda/d)/\lambda^2$ in a way
analogous to (\ref{disorderscale}), resulting in
\begin{equation}
\label{random_vs_elastic}
  \left(
    \frac{\lambda^2}{d\varepsilon_\circ\ln(\pi\lambda/d)}
  \right)^2
  \overline{ 
    \left[
      \nabla^2\varepsilon_{\rm pin}^{_{(i)}} 
    \right]^2
    }
  \simeq
  \Delta \ll 1.
\end{equation}
Hence, the force balance argument leading to the estimate (\ref{vdw})
fully applies in this regime. The van der Waals energy is obtained by
substituting the perturbative result $\overline{u_1^2}\simeq
\xi^2\Delta$ for the correlation function into (\ref{vdw}), yielding
\begin{equation}
  V_{\rm vdw}^{\rm dis}(R) \simeq -\varepsilon_\circ 
  \frac{\Delta}{\kappa^2\,\ln(\pi\lambda/d)}
  \left(\frac{\lambda}{R}\right)^4
  \qquad \mbox{for~}\Delta\ll 1.
\end{equation}

In the opposite limit of strong disorder ($\Delta\gg 1$),
the fluctuations are of order $\overline{u_{\rm pv}^2}\simeq
\xi^2\Delta^{1/2}$ as derived from  the Imry-Ma argument
(\ref{ImryMa}). In this limit, a force equation of the type used above
is not sufficient to determine the response of the pancake position to
dipole-dipole forces due to the presence of a large number of
metastable states. Instead, we have to consider the global energy
minimum. In particular, in the absence of the dipole-dipole
interaction $U_{12}$ (\ref{U_12}) the equilibrium position of each
pancake is determined by the position of the global minimum of the
superposition
\begin{equation}
  E_{el}(\bu_i) + \varepsilon_{\rm pin}^{_{(i)}}(\bu_i),
  \label{Eel_Epin}
\end{equation}
with $E_{el}(\bu_i)$ as defined in (\ref{E_el}). This minimum
will be located at a typical distance of order $\xi\Delta^{1/4}$ from
the vortex center at $\bu_i=0$. Adding now the dipole-dipole energy
$U_{12}$ to 
(\ref{Eel_Epin}), we look for the positional change
$\delta\bu_2$ of pancake 2 induced by the interaction with pancake 1,
keeping $\bu_1$ fixed. The change $\delta\bu_2$ will lead to a
modified dipole-dipole interaction
$U_{12}\rightarrow U_{12}+\delta U_{12}$, from which the vdW interaction can
be derived as $V_{\rm vdw}^{\rm dis} = \overline{\delta U_{12}}/d$, where
the overline again denotes averaging over the disorder.

For the single pancake in stack 2 we are focusing on,
two different scenarios can emerge. First, for large $U_{12}$,
the original position $\bu_2$ may become metastable if the energy
landscape is tilted strongly enough (note that $U_{12}$ is
linear in $\bu_2$) in order that another, formerly local, minimum
becomes the new global minimum; this scenario is illustrated in
Fig.~\ref{landscape}. If,
on the other hand, $U_{12}$ is so weak that the global minimum remains
at the same pinning center, one can again apply perturbation theory to
find an estimate for $\delta\bu_2$.

First, let us have a closer look at the weak $U_{12}$ case. 
Being in the strong disorder limit ($\Delta\gg 1$), the estimate
(\ref{random_vs_elastic}) shows that the restoring forces originating
from the pinning center are now much larger than the elastic
forces. Hence, an additional force ${\bf f}$ applied to pancake 2 will
lead to a response $\bu_2\rightarrow \bu_2+\delta\bu_2$ with 
 $f_\alpha\approx -\partial_\alpha\partial_\beta
\varepsilon_{\rm pin}^{_{(2)}}(\bu_2)\cdot  \delta
u_{2,\beta}$, neglecting the elastic forces.
Substituting the dipole-dipole force (\ref{U_12_force})
for ${\bf f}$, we find a typical displacement $\delta u_2\simeq
(\lambda/R)^2 [1/\Delta^{1/2}\ln(\pi\lambda/d)]
[\overline{u_1^2}]^{1/2}$ which is much 
smaller than $\xi$ for $R\gtrsim\lambda$, and
hence a typical energy change $\overline{\delta U_{12}^{\rm weak}}$ of order
\begin{eqnarray}
  \overline{\delta U_{12}^{\rm weak}} \simeq
  \frac{\varepsilon_\circ d}{R^2}\; \overline{u_1 \cdot\delta u_2}
  \simeq -\varepsilon_\circ d \frac{\lambda^2}{R^4}
  \frac{\Delta^{-1/2}}{\ln(\pi\lambda/d)}\;
  \overline{\bu_1^2}.
\label{deltaUweak}  
\end{eqnarray}
We have neglected orientational dependences here, focusing on amplitudes.
Notice that the factor $\Delta^{-1/2}$ compensates for the 
disorder scaling of the fluctuation square 
$\overline{\bu_1^2}\approx \xi^2\Delta^{1/2}$. Hence, if there were no
pancakes in the stack whose optimal position changes
under the influence of $U_{12}$ (below we will find that this
assumption is actually wrong), this would lead to a weak,
 {\em disorder independent} van der Waals interaction.

\placefigure{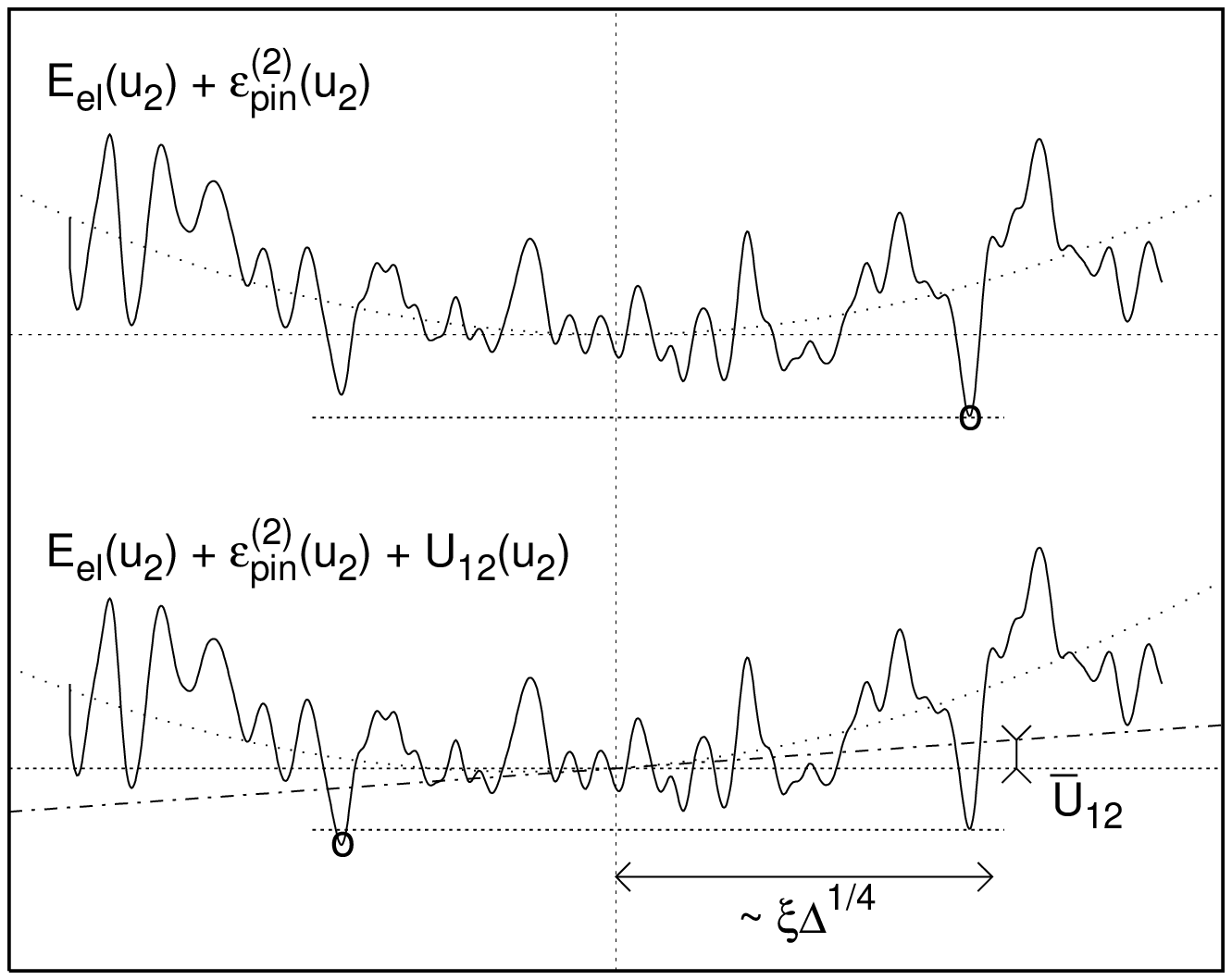}{0.45\textwidth}{In the upper half, a one
  dimensional cut through a typical energy landscape, as given by
  (\protect\ref{Eel_Epin}), is shown; the bare elastic energy is shown
  as a dotted line.  The two deepest minima happen to be located on opposite 
  directions from the origin, in a typical distance of order
  $\overline{u_2^2}\,^{1/2}\simeq \xi\Delta^{1/4}$.  In the lower
  half, a linear energy contribution $U_{12}(u_2)$ is added, sketched
  as a dashed-dotted line. Note that the center of the new parabola
  $E_{el}(u_2)+U_{12}(u_2)$ has been shifted to the left hand
  site. The addition of $U_{12}$ changes the order of the 
  minima, hence this scenario corresponds to the 'strong $U_{12}$'
  case; the deepest minima are marked by small circles.}

The second possible scenario emerges when 
$U_{12}$ is strong enough to induce a change of
the order of the deepest energy minima for the pancake in stack 2. In
this case, it will induce a shift $\delta 
u_2$ of order $\xi\Delta^{1/4} \gg \xi$, since the new minimum may be
located anywhere in a distance $(\overline{\bu_2^2})^{1/2}$ from the
center. Hence,
\begin{equation}
  \overline{\delta U_{12}^{\rm strong}}\approx-\overline{U_{12}},
\label{deltaUstrong}
\end{equation}
where $\overline{U_{12}}$ is taken 
to be a typical dipole-dipole energy at the distance $R$,
\begin{equation}
\label{typicalGain}
   \overline{U_{12}\rule{0ex}{1.5ex}} \equiv 
   U_{12}\left(|\bu_2|\approx \xi\Delta^{1/4}\right) \;\approx\;
  \frac{2\varepsilon_\circ d}{R^2} \xi^2 \Delta^{1/2}.
\end{equation}

So far, we have focused on a single pair of pancake vortices in the
same layer. Now,  a vortex line consists of a whole stack of pancake
vortices. In a certain fraction $\eta_\Delta$ of these, the {\em
  strong} $U_{12}$ scenario will apply, where the
dipole-dipole interaction induces a positional change
$\delta\bu_2\gg\xi$; the rest will stay within its old minimal energy
position, giving only a marginal contribution to the vdW
interaction. By estimating this fraction $\eta_\Delta$ via level
statistics of the Gaussian random distribution, we can derive an
estimate for the strength of the vdW energy as
\begin{equation}
  V_{\rm vdw}^{\rm dis}=\frac{1}{d}
  \left(
    \eta_\Delta \overline{\delta U_{12}^{\rm strong}}
    + (1-\eta_\Delta) \overline{\delta U_{12}^{\rm weak}}
  \right) \, .
\label{vdwFromEta}
\end{equation}
Below, we will find that the second term in (\ref{vdwFromEta}),
emerging from those
pancakes with $\delta u_2\ll \xi$, can be neglected.

We will now calculate the fraction $\eta_\Delta$ of 'active' pancakes
in the stack. First, consider once again the situation
in the absence of dipole-dipole forces, where the two pancakes take
their optimal positions $\bu_1$ and $\bu_2$. The dipole-dipole
interaction will induce a force $-(2\varepsilon_\circ d/R^2)\,{\bf b}$
on pancake $\bu_2$, with ${\bf b}=\bu_1-2({\bf \hat
  n}\cdot\bu_1)\hat{\bf n}$ (note that ${\bf b}^2=\bu_1^2$). Let us
divide the area of size $\xi^2\Delta^{1/2}$ which is accessible to 
$\bu_2$ in two half circles, the dividing line being
perpendicular to ${\bf b}$. In half of the cases, the deepest minimum
will be located 
({\em before} the addition of the dipole-dipole force)
in the half space where the energy landscape is
shifted upwards, as shown in Fig.~\ref{landscape}. These pancakes
can probably profit from $U_{12}$ by taking a 
new optimal position in the other half space.
With the disorder correlation length given by $\xi$, 
the energy landscape that is accessible to the pancake can be
represented by 
\begin{equation} 
\label{independentRealizations}
  N\simeq\Delta^{1/2}
\end{equation} 
effectively independent disorder realizations. We can hence
represent the energy (\ref{Eel_Epin}) in each half circle by a set of $N/2$  
Gaussian random numbers $\varepsilon_m^i$ ($i=1,2$ for the two sets,
$m=1,\ldots,N/2$), each measuring the 
minimum of the total energy in a cell of size $\xi^2$.

In Appendix \ref{AppLevelStat}, the
probability density $p_{12}(\delta \varepsilon)$ for the
difference $\delta\varepsilon=\varepsilon_{\rm
  min}^{1}-\varepsilon_{\rm min}^{2}$ between the smallest values of
the two sets, 
respectively, is derived; cf.\ Eq.~(\ref{p12_def}). Given this
probability density, the fraction of 'active' pancakes can be 
estimated as 
\begin{equation}
\label{integral_p12}
  \eta_\Delta \approx \int_0^{\overline{U_{12}}} d(\delta\varepsilon)
  \; p_{12}(\delta\varepsilon) \; .
\end{equation}
This integral gives the probability that the energy difference between
the deepest minima of the two sets is positive and smaller than the
typical energy gain $\overline{U_{12}}$.

For small enough values $\overline{U_{12}}$, corresponding 
to distances $R$ between the two vortices 
larger than some distance $R_c$ which we quantify below, we
can approximate the probability density $p_{12}(\delta\varepsilon)$ in
the integrand of Eq.\ (\ref{integral_p12}) by its value at
$\delta\varepsilon=0$,  
\begin{equation}
  \eta_\Delta \;\approx\;  p_{12}(0)\,\overline{U_{12}}  \;\approx\;
  \frac{\ln^{1/2} (N/2)}{\sqrt{2\pi}} \,\frac{\overline{U_{12}}}
  {\bar\varepsilon} ,
\label{eta_Delta}
\end{equation}
where we have used (\ref{p12_saddlepoint}), and $\bar\varepsilon$ is
defined as 
\begin{equation}
\label{eps_variance}
  \bar\varepsilon = 
  \varepsilon_\circ d \,\frac{\ln(\pi\lambda/d)}{\kappa^2} \,\Delta^{1/2}.
\end{equation}
Now, plugging the results from
Eqs.~(\ref{eta_Delta}) and (\ref{deltaUstrong}) into (\ref{vdwFromEta}), we
find 
\begin{eqnarray}
  V_{\rm vdw}^{\rm dis}(R\gtrsim R_c)  
  \simeq -\varepsilon_\circ\,\frac{4}{\sqrt{2\pi}}
  \,\frac{\Delta^{1/2}\,\ln^{1/2}(\Delta^{1/2})}{\kappa^2\,\ln(\pi\lambda/d)} 
  \,\left(\frac{\lambda}{R}\right)^4\nonumber\\
  \mbox{for~}\Delta \gg 1.\quad \rule{0cm}{3ex}
\label{vdwdis_largeR}
\end{eqnarray}
It is easy to check that the second contribution in (\ref{vdwFromEta}) is
smaller than the first one by a factor of $\Delta^{1/2}$ and can hence be
neglected. 
Re-substituting a factor of $\xi^2\Delta^{1/2}$ by $\overline{\bu_1^2}$ in
(\ref{vdwdis_largeR}), we have finally arrived at an expression for
the van der Waals energy which is identical to (\ref{vdw}) up to a
factor of ${\cal O}(1)$ and a subdominant logarithmic contribution in
the disorder strength $\Delta$. It is consistent with the result from 
the systematic derivation obtained in section \ref{sec_statmech}, again
up to a numerical factor and the $\ln(\Delta^{1/2})$ term.

Finally, we note that for distances $R$ smaller than $R_c$,
$\eta_\Delta\rightarrow 1$, leading, via (\ref{typicalGain}), to the
saturation value 
\begin{equation}
\label{vdwdis_smallR}
  V_{\rm vdw}^{\rm dis}(R\lesssim R_c) \simeq
  -\varepsilon_\circ \frac{\Delta^{1/2}}{\kappa^2}\,
  \left(\frac{\lambda}{R}\right)^2.
\end{equation}
At distances $R<R_c$, the vdW attraction is however dominated by the
direct repulsion.
The expressions (\ref{vdwdis_largeR}) and  (\ref{vdwdis_smallR}) match
at $R\approx R_c$, where the crossover distance is given by
\begin{equation}
\label{criterion_R}
  R_c^2 \simeq
  \frac{4\ln^{1/2}\left( \Delta/4\pi \right)}
  {\ln(\pi\lambda/d)} \,\lambda^2 .
\end{equation}
$R_c$ has a simple meaning:
For $R\approx R_c$, the typical dipole-dipole energy
$\overline{U_{12}}$ is of the same order as the typical energy
difference  between the two deepest minima from the two sets.
Since the  prefactor on the right hand side of (\ref{criterion_R})
depends only logarithmically on $\Delta$ and $\lambda/d$, $R_c$ will
be of order $\lambda$, so that expression (\ref{vdwdis_largeR})
correctly predicts the vdW interaction for all
relevant values $R\gtrsim \lambda$.

With the parameters for the different length scales characteristic for
BiSCCO as cited above, together with the disorder strength
$\Delta=\Delta(\pi\lambda/d)\approx 300$, $N=\sqrt\Delta$ is as low as
$\approx 20$. With this value for $\Delta$, we find $R_c\approx \tilde
c \lambda$ with $\tilde c\approx 1.1$.

\subsection{Statistical mechanical approach}
\label{sec_statmech}

The vdW attraction can be derived using  a more formal statistical
mechanical approach where we consider the free energy of two FLs
fluctuating due to thermal energy and impurities. In the absence of
disorder we obtain the vdW attraction due to the temperature induced
fluctuations of the lines. 

Our starting point is the free energy in (\ref{iso}) that contains a
self energy part ${\cal F}_0$ and a mutual interaction part 
${\cal F}_{\rm int}$ consisting of terms with $i\neq j$. To proceed
further it is convenient to write the mutual interaction explicitly in
terms of components:
\begin{eqnarray}
  \label{comp}
  &&{\cal F}_{\rm int}=\frac{\Phi_0^2}{4 \pi}
  \int dk\,dz_1dz_2\, e^{ik(z_1-z_2)} \times \\
  && \qquad\quad 
  \Big[\,V_{zz}^{\rm int}(\bR+\bu_1(z_1)-\bu_2(z_2),k)+  \nonumber\\
  &&\qquad\qquad t_{1\alpha}(z_1)t_{2\beta}(z_2)\,
  V_{\alpha \beta}^{\rm int}({\bf  R}+\bu_1(z_1)-\bu_2(z_2),k)\,\Big]
  \nonumber 
\end{eqnarray}
where we have used $d{\bf s}_\mu=({\bf t}(z),1)\,dz$ with
${\bf t}(z)=\partial_z\bu(z)$.
Here and below, each integral over $k$ implies a factor of $1/2\pi$.
We split ${\cal F}_{\rm int}$ into a longitudinal 
part ${\cal F}_\parallel$ with the term $V_{zz}$ and a transverse
part  ${\cal F}_\perp$  with the term proportional to $t_{1\alpha}(z_1)
t_{2\beta}(z_2)$. 
The partition
function which is a path integral corresponding to the weighted sum
over all possible displacements is 
\begin{equation}
  \label{partition}
  Z(R)=\int {\cal D}[{\bf u}_1(z)]\;{\cal D}[{\bf u}_2(z)] \;\exp[-{\cal
    F}(R)/T]. 
\end{equation}
In the presence of impurities, the disorder averaged free energy 
can be written as 
\begin{equation}
  \label{free}
  F(R)=-T \,{\overline {\ln Z(R)}}=-T\,{\overline {\ln \langle \exp[-{\cal
      F}_{\rm int}(R)/T]\rangle_0}},
\end{equation}
where the average $\langle ...\rangle_0$ is taken with respect to 
the self energy part ${\cal F}_0$. 
The  effective interaction between the FLs follows immediately from
the cumulant expansion 
\begin{equation}
  \label{effint}
  LV_{\rm eff}(R)\approx {\overline {\langle {\cal F}_{\rm int}\rangle}}
  - {\overline {[\langle {\cal F}_{\rm int}^2\rangle_0 - \langle {\cal
      F}_{\rm int} \rangle_0^2]}}/2T
\end{equation}
 Our aim at this point is 
to look for a long range vdW type
attraction  in $V_{\rm eff}$. 

\subsubsection{First cumulant}
It is expected that the fluctuation induced vdW attraction should
originate from the second cumulant. We first show that  no such
contribution arises from the first order term in (\ref{effint}). 

Starting with the longitudinal part
\begin{eqnarray}
  \label{1stord}
  {\cal F}_\parallel &=& \frac{\Phi_0^2}{4\pi} 
  \int dk \,dz_1\,dz_2 \;e^{i k(z_1-z_2)} \times\nonumber\\ 
  && \qquad\;  V_{zz}^{\rm int}(\bR+\bu_1(z_1)-\bu_2(z_2), k) 
\end{eqnarray}
a straightforward approach is to expand the interaction term
$V_{zz}^{\rm int}(\bR+\bu_1(z_1)-\bu_2(z_2), k)$ in ${\bf u}$. 
The O(1) term 
\begin{equation}
  \label{longi}
  \frac{\Phi_0^2}{4\pi} \int dz_2 \; V_{zz}^{\rm int}(\bR, k=0)
\end{equation}
with $V_{zz}^{\rm int}(\bR, k)$ as given by (\ref{V_zz}),
leads to a short range potential decaying exponentially for
$R>\lambda$. The higher order terms lead to higher order derivatives
in $\bR$, as can be seen by inspecting the O($u^2$) term 
\begin{eqnarray}
  &&\frac{\Phi_0^2}{4\pi} \int \!\! dk dz_1 dz_2
  \sum_{\alpha \beta} 
    \frac{\partial}{\partial R_\alpha} 
    \frac{\partial}{\partial R_\beta} \,
    V_{zz}^{\rm int}(\bR,k) 
  \;e^{ik(z_1-z_2)}\times \nonumber \\
  && \qquad
  \overline{
    \big\langle(u_{1\alpha}(z_1)-u_{2\alpha}(z_2))
    (u_{1\beta}(z_1)-u_{2\beta}(z_2))\big\rangle
    },
\end{eqnarray}
and hence again are of short range in nature.

The $O(1)$ term from a similar expansion of the exponential of the
transverse part  
\begin{equation}
  \label{trans}
  \frac{\Phi_0^2}{4\pi} \int dk \,dz_1\, dz_2\ 
  {\overline{\langle t_{1\alpha}(z_1)t_{2\beta}(z_2)\rangle}}
  \,V^{\rm int}_{\alpha\beta}(\bR,k) \,e^{ik(z_1-z_2)}
\end{equation}
requires averaging over the impurity distribution for the flux line
displacements.  We approximate the averaging as 
${\overline{\langle {t}_{1\alpha}(z_1){t}_{2\beta}(z_2)
\rangle}}\approx 
{\overline{\langle {t}_{1\alpha}(z_1)\rangle}}\ \ {\overline{\langle {
      t}_{2\beta}(z_2)\rangle}}$. Such a decomposition makes sense only
when the flux lines are far apart so that the preferable impurity
sites of line $1$ do not affect the configuration of the line $2$. 
With this approximation it is now evident that the transverse part
vanishes since  ${\overline{\langle {t}_{1\alpha}\rangle}}=0$.
The $O(u)$ term contribution vanishes for the same reason.
$O(u^2)$ 
terms require evaluation of several correlation 
functions. As an example we pick up one term and show that it vanishes
due to the symmetry of the integral. In the same way, it can be shown
that also the other terms vanish. Let us consider the term 
\begin{eqnarray}
  \label{usquare}
  2\frac{\phi_0^2}{4\pi}\int dk\,dz_1\,dz_2\; 
  \frac{\partial^2}{\partial R_x^2} 
    V^{\rm int}_{\alpha\beta}(\bR,k) \;
  e^{ik(z_1-z_2)}\times 
  \nonumber\\ 
{\overline{\langle {t}_{1\alpha}(z_1){ t}_{2\beta}(z_2)
{u}_{1x}(z_1){u}_{2x}(z_2)\rangle}}  .
\end{eqnarray}
As before we assume that the averaging over the impurity distribution
for FL $1$ and $2$ is uncorrelated in nature. 
The above correlation function is therefore equivalent to 
\begin{eqnarray}
  \label{fourpt}
 && {\overline{\langle t_{1\alpha}(z_1) u_{1x}(z_1)\rangle}}\;\,
  {\overline{\langle t_{2\beta}(z_2) u_{2x}(z_2)\rangle}} \\
 && =-\int d\{k_{i}\}k_{1}k_{3}\;
 \big(
   C_T^{\alpha x}(k_{1},k_{2})+C_{\rm dis}^{\alpha
     x}(k_{1},k_{2})
   \big) 
   \times\nonumber\\
   && \big(
     C_T^{\beta x}(k_{3},k_{4})+C_{\rm dis}^{\beta
       x}(k_{3},k_{4})  
   \big)
   e^{i(k_{1}+k_{2})z_1} e^{i(k_{3}+k_{4})z_2}, \nonumber
\end{eqnarray}
where we have used 
$C_{\rm dis}^{\alpha\beta}(k_1,k_2)={\overline{\langle {
      u}_{1\alpha}(k_1)\rangle \langle {u}_{1\beta}(k_2)\rangle}}$ 
and
$C_T^{\alpha\beta}(k_1,k_2)=
{\overline{\langle{u}_{1\alpha}(k_1){u}_{1\beta}
    (k_2)\rangle}} -{\overline{\langle {u}_{1\alpha}(k_1)\rangle
    \langle {u}_{1\beta}(k_2)\rangle}}$
in a straightforward generalization of the corresponding quantities
(\ref{C_T}) and (\ref{C_dis}) defined in section \ref{ChainSection}.
Due to translational invariance, 
both $C_{\rm dis}^{\alpha\beta}(k_1,k_2)$ and 
$C_T^{\alpha\beta}(k_1,k_2)$ are diagonal in their arguments,
i.e., they include a factor of $\delta(k_1+k_2)$.
Using this property of the correlation functions, the integration over
$z_1$, $z_2$ and $k$ in (\ref{usquare}) leads to 
\begin{eqnarray}
  \label{fourpt1}
  -2\frac{\Phi_0^2}{4\pi}\int dk_{1} dk_{3}\;\, k_{1}k_{3}\,
  \frac{\partial^2}{\partial R_x^2} 
    V_{\alpha\beta}^{\rm int}({\bf R},k=0) \,\times\nonumber\\
  \big(C_T^{\alpha x}(k_{1}) +C_{\rm dis}^{\alpha x}(k_{1})\big)
  \big(C_T^{\beta x}(k_{3}) +C_{\rm dis}^{\beta x}(k_{3})\big),
\end{eqnarray}
where we have switched back to the notation as in
(\ref{C_T}) and (\ref{C_dis}). Now, both $C_{\rm dis}(k)$ and
$C_T(k)$ are symmetric in $k$, which leads to an integrand
odd in $k_{1}$ and $k_{3}$. The integration regime given by
$k_{1},$ $k_{3}=-\pi/d\ldots\pi/d$, the integral hence vanishes by
symmetry.

\subsubsection{Second cumulant}
Next we go over to the second term in (\ref{effint}). The pure
longitudinal part ${\cal F}_\parallel^2$ again leads to only a short
range interaction which has the same effect as before  of
renormalizing the   short range potential between the flux lines.
The cross term of the longitudinal and the transverse parts contains
${\overline{\langle {t}_{1\alpha}{t}_{2\beta}\rangle}}$ and
therefore vanishes after disorder averaging if the flux lines are
assumed to be far apart. 
We are left with the contribution from the pure transverse part
${\overline {[\langle {\cal F}_{\perp}^2\rangle_0 - \langle {\cal
      F}_{\perp} \rangle_0^2]}}/2T$, which has to be analyzed in a
more detail. Indeed, as we will show 
the contribution to the vdW attraction per unit length can be found as
\begin{equation}
  \label{statvdw}
  V_{\rm vdw}=-\frac{1}{2TL}
  {\overline {[\langle {\cal F}_{\perp}^2\rangle_0 - 
      \langle {\cal F}_{\perp} \rangle_0^2]}}\;.
\end{equation}
To lowest order in the displacements, ${\cal F}_\perp$ reads 
\begin{equation}
  \label{zfour}
  {\cal F}_\perp=\frac{\Phi_0^2}{4\pi}\int dk\;k^2 \;
  u_{1\alpha}(-k)u_{2\beta}(k) \,V^{\rm int}_{\alpha\beta}(\bR,k)\;, 
\end{equation}
with $V^{\rm int}_{\alpha\beta}(\bR,k)$ as defined in
(\ref{V_xx}) or (\ref{V_xx_eps0}) for $\varepsilon>0$
or $\varepsilon\rightarrow 0$, respectively.

For the evaluation of (\ref{statvdw}), it is now necessary to evaluate
the correlation function 
\begin{eqnarray}
C(k)&=&\overline{\langle
  u_{1\alpha}(-k)u_{2\beta}(k)u_{1\gamma}(-k')u_{2\delta}(k')\rangle} 
\nonumber\\
&&-  
 \overline{\langle u_{1\alpha}(-k)u_{2\beta}(k)\rangle \langle
 u_{1\gamma}(-k')u_{2\delta}(k')\rangle}.
\end{eqnarray}
Using the above decomposition for disorder averaging, this can
be rewritten as 
\begin{eqnarray}
  \label{rewrit}
  C(k)&=&
  C_T^{\alpha \gamma}(-k,-k') 
  C_T^{\beta \delta}(k,k')+C_{\rm dis}^{\alpha
    \gamma}(-k,-k')\times \nonumber\\
&& C_T^{\beta
  \delta}(k,k')+C_{\rm dis}^{\beta 
  \delta}(k,k')C_T^{\alpha \gamma}(-k,-k').
\end{eqnarray}
The first term in (\ref{rewrit}) involving only $C_T$ leads to
the  vdW attraction found in the pure system. This follows from the
fact discussed in section \ref{ChainSection} that the disorder
does not affect $C_T$. Substituting this term in
(\ref{statvdw}), we find  
\begin{eqnarray}
  \label{vdwi}
  V_{\rm vdw}^{\rm th}&=&\sum_{\alpha \beta \gamma \delta}
  \int dk\,dk'\; k^2\,k'^2\; 
  V^{\rm int}_{\alpha\beta}(\bR,k)\;
  V^{\rm int}_{\gamma\delta}(\bR,k') \times \nonumber \\
  && \quad\qquad C_T^{\alpha \gamma}(-k,-k') \,  
  C_T^{\beta\delta}(k,k')
\end{eqnarray} 
Using the fact that $C_T^{\alpha \gamma}$ and $V^{\rm
  int}_{\alpha \beta}$ are diagonal in components, and substituting  
$V_{\alpha \beta}^{\rm int}$ from (\ref{V_xx_eps0}), we find in the
extremely decoupled limit 
\begin{eqnarray}
  \label{vdwj}
  V^{\rm th}_{\rm vdw}&=&\frac{1}{2T}\frac{\Phi_0^4}{(4\pi)^2}
\frac{1}{(2\pi R^2)^2} \int dk \frac{k^4}{(1+\lambda^2
    k^2)^2} \left(\frac{T}{k^2\varepsilon_l(k)}\right)^2\nonumber\\
&\simeq&-\frac{\varepsilon_\circ}{\ln^2(\pi \lambda/d)}
\frac{T}{d\varepsilon_\circ}
 \left(\frac{\lambda}{R}\right)^4
\end{eqnarray}
In the last step, we have approximated the integrand by its value at
the upper integration limit $\pi/d$ where the contribution to the
integral is the largest. 

In the continuous anisotropic case one has to consider the 
more general form of the potential in (\ref{V_xx}).
 $K_1(x)$ has the asymptotic properties 
$K_1(x\ll 1)\simeq 1/x$ and $K_1(x\gg 1)\sim
e^{-x}/\sqrt{x}$. Due to the exponential decay of $K_1(x)$ for large 
arguments, 
the vdW attraction is exponentially suppressed for
$R>\lambda/\varepsilon$. For smaller $R$, 
without much loss we may  set the upper limit of the integral to
$1/\varepsilon R$ and use $K_1(x)\sim 1/x$ in the integrand. Again the
contribution is maximum at the upper limit.  Thus we find the
thermal vdW attraction in the continuous anisotropic case 
\begin{equation}
  \label{vdwk}
  V^{\rm th}_{\rm vdw} \simeq -\frac{\varepsilon_\circ}{\ln^2(\pi
    \lambda/\varepsilon R)} 
  \frac{T}{\varepsilon \lambda
    \varepsilon_\circ}\left(\frac{\lambda}{R}\right)^5
  \quad\mbox{for~} \frac{d}{\varepsilon} < R < \frac{\lambda}{\varepsilon}.
\end{equation}
For $R<d/\varepsilon$, $K_1(x)\simeq 1/x$ in the whole
integration range, so that (\ref{vdwj}) applies in this regime.

As expected, the vdW attraction in a pure system vanishes as
$T\rightarrow 0$. Expressions (\ref{vdwj}) and (\ref{vdwk})
for the thermally induced vdW attraction have been obtained
previously in Ref.~\cite{BlatterGeshkenbein}. 

Using the fact that the correlation functions are diagonal in
components, the most general form of the vdW attraction
in the presence of the disorder and at $T\neq 0$ is
\begin{eqnarray}
  \label{vdwdis}
  V^{\rm dis}_{\rm vdw}&=&
  -\frac{1}{4T}\frac{\Phi_0^4}{(4\pi)^2}\int_0^{\pi/d} dk \,k^4 
  \;[\,V_{xx}^{\rm int}(\bR,k)\,]^2 \times\nonumber\\
  && \qquad\qquad(C_T^2(k)+2 C_T(k)C_{\rm dis}(k)).
\end{eqnarray}
In order to see purely the effect of the disorder, one has to study the
the term with $C_{\rm dis}$ in this equation. In section
(\ref{sec:CorrFunc}), we have derived expressions for $C_{\rm dis}(k)$
both in the limit $\lambda k\ll 1$ and $\lambda k\gg 1$; the exact
behavior at $\lambda k$ of order unity is however
unclear. Nevertheless, provided the roughness exponent $\zeta<1$, one finds
that the dominant contribution to the $k$-integral comes from large
wave vectors $k\lesssim\pi/d$. Hence, the vdW attraction can be
derived by substituting the expression for $C_{\rm dis}$ from
(\ref{cdiss}), and proceeding as in the thermal case. It is found to
have the same $R$-dependence as in the thermal case both in the extreme 
decoupled and in the continuous case, respectively. We give the result
in form of an ad-hoc interpolation formula which subsumes both limits $R\ll
d/\varepsilon$ and $R\gg d/\varepsilon$, 
\begin{equation}
  \label{vdwdis1}
  V_{\rm vdw}^{\rm dis}\simeq
  -\frac{\varepsilon_\circ}{\kappa^2}
  \left(\Delta_0\frac{\lambda}{d}\right)^\eta\left(\frac{\lambda}{R}\right)^4
\frac{d}{d+\varepsilon R}\left(\ln\frac{\pi \lambda}{d+\varepsilon
  R}\right)^{-1-2\eta},
\end{equation}
with $\Delta_0$ and $\eta$ as defined in (\ref{delta}) and 
(\ref{eta}), respectively. Like in the thermal case, (\ref{vdwdis1})
holds only for $R<\lambda/\varepsilon$; for larger $R$, the vdW
attraction is exponentially suppressed. Note that in the decoupled limit
$\varepsilon\rightarrow 0$, the result from the intuitive approach
(\ref{vdwdis_largeR}) is reproduced up to numerical factors.

Finally, we note that higher order terms in the
displacement $\bu$ in the expansion of the interaction potential 
in (\ref{comp}) introduce derivatives of $V_{\alpha\beta}^{\rm int}$
by components of $\bR$ in (\ref{statvdw}). The ${\cal O}(\bu^2)$ 
term, for example, yields a finite contribution at $T=0$ and finite
disorder, but vanishes $\sim R^{-6}$ or $R^{-7}$, respectively, and
can hence be neglected.  

\section{Phase Diagram}
\label{phasediag_sec}

In the low-field limit, where the magnetic induction $B$ is of order
a few Gauss, the attractive van der Waals interaction between vortex
lines has consequences for the phase diagram of anisotropic
superconductors. With the external magnetic field $H$ fixed, one has
to analyze the Gibbs free energy density  
\begin{equation}
\label{GibbsGeneral}
  G(\ell;H,T,T_{\rm dis})=F(\ell;T,T_{\rm dis})
  -\frac{\varepsilon_\circ\ln\kappa}{\ell^2}\,\tilde h 
\end{equation}
which has to be minimized with respect to the mean FL distance $\ell$.
$\tilde h=(H-H_{c_1}^0)/H_{c_1}^0$ is the reduced magnetic
field, $H_{c_1}^0=4\pi\varepsilon_\circ\ln\kappa/\Phi_0$ being the
unrenormalized
lower critical field, and $F(\ell;T,T_{\rm dis})$ is the free
energy density. $\ell$ is related to the
magnetic flux via $B=\Phi_0/\ell^2$.

In section \ref{subsec:thermal}, we will restrict ourselves to the
pure thermal case $T>0$ with vanishing disorder $(T_{\rm dis}=0)$. We
will give a short overview over approaches applied previously to the
determination of the effective Gibbs free energy density, and review a
scaling approach in the first subsection \ref{sec_scaling}. In the
following subsections, we will go beyond scaling arguments and apply
functional renormalization to the FL interaction potential. This will
enable us to make quantitative predictions for the phase diagram.

Finally, we will briefly consider the case with disorder in section
\ref{subsec:disorder}, restricting ourselves again to a scaling approach.

\subsection{Pure case $T_{\rm dis}=0$.}
\label{subsec:thermal}

In the absence of the vdW interaction, the free energy density $F(\ell,T)$
can be represented by the superposition of the bare interaction energy
$z\varepsilon_\circ K_0(\ell/\lambda)/\ell^2$ and an entropic
contribution $\sim (T/\ell^2)/\ell^2$. Here, $z$ is the lattice
coordination number ($z=6$ for a triangular lattice). The entropic
term describes the reduction of the 
vortex line entropy due to its confinement within the cage set up by
its neighbors \cite{Nelson}. It can be viewed as a renormalization of
the bare interaction, taking into account (thermal) fluctuations on
length scales between $\lambda$ and $\ell$.
While this renormalization of a short range, purely repulsive function
can be handled within a one parameter renormalization group
calculation \cite{NelsonSeung}, it is not simply possible to account
for the attractive tail of the van der Waals interaction within this
approach. 

In recent studies, different methods have been applied to account for
the van der Waals interaction in the Gibbs free energy density. In the
original work \cite{BlatterGeshkenbein}, the bare vdW energy,
evaluated at the mean FL distance $\ell$,  was
added to the free energy. This means that only contributions
from the vdW energy on the length scale $\ell$ are taken into account,
which leads to a gross underestimation of its influence because of its
rapid decay for $R>\lambda$. In \cite{NattermannMukherji}, on the
other hand, only the much more important contribution on the scale
$R_{\rm min}$ has been taken into account, where $R_{\rm
  min}$ is the position of the global minimum of the bare potential
(\ref{bareV0}), see below. This method, which makes use of simple
scaling arguments, will be reviewed in the following section.
An explicit integration of the fluctuations can be
performed using a functional renormalization group (RG)
calculation. The description of this procedure and its results
is the main focus of this section. The
results from the RG will be used to check whether the scaling
arguments lead to a valid description of the effective free energy,
and to quantify numerical factors that cannot be specified within this
approach.

The list of alternative methods is completed by an approach that
makes use of the mapping of the vortex line problem onto $2d$-Bosons
\cite{bg2}, and a variational procedure. This latter approach will be
considered in a forthcoming publication \cite{Schwartz}. 

\subsubsection{Scaling approach}
\label{sec_scaling}

The superposition of the bare repulsion and the attractive vdW
interaction between flux 
lines, 
\begin{equation}
\label{bareV0}
  V_0(R) \;\equiv\; 2\varepsilon_\circ K_0(R/\lambda) +
  V_{\rm vdw}(R),
\end{equation}
results in a minimum of their interaction energy at a
distance $R_{\rm min}\approx \alpha \lambda$ $(\ll \ell)$ (the typical
form of the bare interaction potential is shown in the inset of
Fig.~\ref{bare_pot}). 

The position of the minimum can be conveniently quantified by
considering the function $v_n(x)=K_0(x)-a_{\rm vdw}/x^n$, where
$n$=4 or 5. In Fig.~\ref{bare_pot}, $v_4(x)$ is shown for  some values
of $a_{\rm vdw}$. 
For a huge amplitude range $10^{-10}<a_{\rm vdw}<1$, we
find from a numerical fit that the 
minimum position $\alpha_n$ of $v_n(x)$ is well described by the
functional form $\alpha_n\simeq \tilde x_n + 1.95
\ln^{0.87}(1/a_{\rm vdw})$ with $\tilde x_4=8.5$ and $\tilde
x_5$=12.2.
From the prefactors of the thermal vdW interaction as given in
(\ref{vdwj}) and (\ref{vdwk}), using  the parameters for BiSCCO, we
read off that the dimensionless vdW amplitude $a_{\rm vdw}$ takes
values of order $2\cdot 10^{-5}\,T/$K for $n=4$ and $10^{-4}\,T/$K for
$n=5$. Using the latter value, this corresponds to the minimum
position ranging from $\alpha\approx 19$ at 
$T=100\,$K to $\alpha\approx 26$ at $T=1\,$K.
Also, the width of the minimum only weakly depends on temperature,
being of the order of $\beta \lambda$ with $\beta\approx 10$. Like
$\alpha$, $\beta$ becomes larger for smaller $T$.

\placefigure{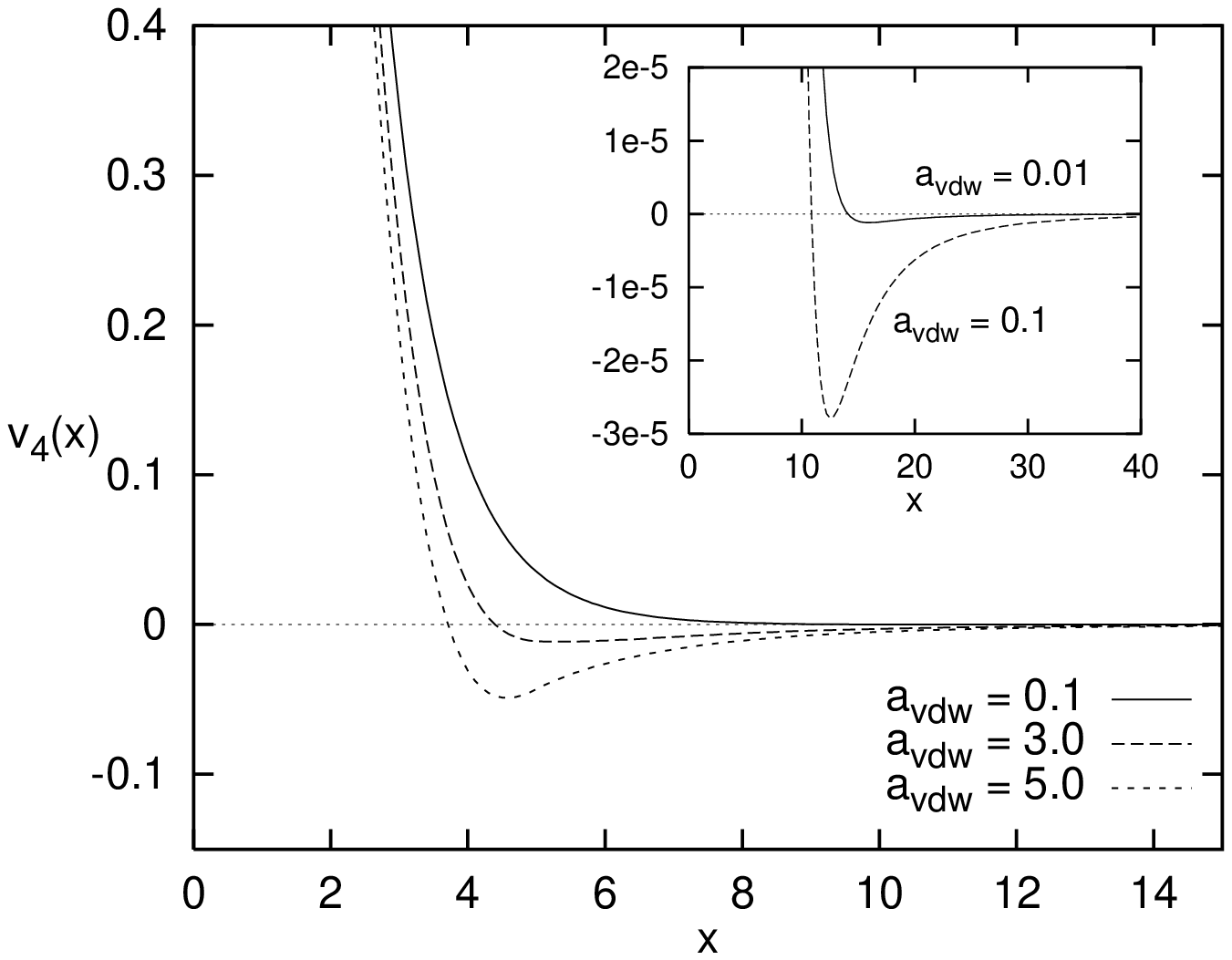}{0.49\textwidth}{For some values of the vdW
  amplitude $a_{\rm vdw}$, the (dimensionless) bare interaction
  potential $v_4(x)=K_0(x)-a_{\rm vdw}/x^4$ is shown, with a smooth
  cutoff of the algebraic part between $x=1$ and $x=5$. The exact form of the
  cutoff will be specified below, see Eqs.~(\protect\ref{bareV0spec})
  and (\protect\ref{CutOff}).  The inset shows a
  magnified view for two smaller amplitudes $a_{\rm vdw}$. }

Due to the strong distance dependence of the vdW attraction, its main
contribution comes from those configurations where the line pair is at
a distance $R_{\rm min}$.  In the estimate of this contribution
developed below, we will account for both the thermal and the disorder
dominated case, which are characterized by the length scales $L_T$ and
$L_{\rm dis}$, respectively.

We estimate the average vdW interaction by considering the
configurations of a single line in the absence of any FL interaction.
With $u\!\approx \!  \lambda (L/L_{{\rm T},{\rm dis}})^{\zeta}$ for
the displacement of a single FL (with the roughness exponent $\zeta$
as discussed in section \ref{sec:SingleFLHamiltonian}), we find from
$u\approx \ell$ for the mean distance $L_\parallel$ between two line
segments reaching a minimum $L_\parallel \approx L_{\rm
T,dis}(\ell/\lambda)^{1/\zeta}$.  The length $L_s$ of the segment over
which the line stays in the minimum follows from the same argument as
$L_s\approx L_{\rm T,dis}\ \beta^{1/\zeta}$. Thus, the contribution
from the vdW attraction to the Gibbs free energy density
(\ref{GibbsGeneral}) is of the order \cite{NattermannMukherji}
\begin{equation}
\label{ScalingVDW}
  \frac{1}{\ell^2} V_{\rm vdw}(R_{\rm min})\frac{L_s}{L_\parallel}
  \approx V_{\rm vdw}(R_{\rm min})
  \left(\frac{\lambda}{\ell}\right)^{2+1/\zeta}
  \frac{\beta^{1/\zeta}}{\lambda^2},
\end{equation}
which is much larger than the vdW interaction at the mean distance $\ell$.
For  thermal fluctuations $(\zeta=1/2)$ the mean vdW attraction has
therefore the same $\ell$ dependence as the entropic repulsion. In this
case, the  result can also be obtained by mapping the problem onto
$2d$- Bosons \cite{bg2}.

The Gibbs free energy density $G(x;H,T,T_{\rm dis}=0)$
can hence be written in the following form \cite{NattermannMukherji} 
\begin{equation}
  G(x;H,T,0)\approx
  \frac{\varepsilon_\circ}{\lambda^2x^2}
  \left\{
    zK_0(x)+\frac{\gamma_T-\delta_T}{x^2} - \tilde{h}\ln{\kappa}
  \right\}
\label{GibbsScaling}
\end{equation}
which has to be minimized with respect to $x=\ell/\lambda$.
Expression (\ref{GibbsScaling}) has to be considered as an interpolation
between the regimes dominated by the bare interaction at high $B$
and the different fluctuation induced interactions at low $B$, 
respectively.
The strength of the entropic repulsion is given by
\cite{BlatterGeshkenbein} 
\begin{equation}
\label{gamma_T}
  \gamma_T\approx 9.08
  \left(T/\varepsilon_\circ\lambda\right)^2,
\end{equation} 
and the prefactor of the term following from the thermally induced vdW
interaction is  
\begin{equation}
\label{delta_T}
  \delta_T\approx 
  \left\{ 
    \begin{array}{cl}
      \ds c_T
      \,\frac{\beta^2}{\alpha^4}\,
      \frac{T}{\varepsilon_\circ d\,\ln^2(\pi\lambda/d)} &
      \mbox{ for}\quad  R_{\rm min} < d/\varepsilon \\
      \ds c'_T
      \,\frac{\beta^2}{\alpha^5}\,
      \frac{T}{\varepsilon\varepsilon_\circ \lambda \,
        \ln^2(\pi/\varepsilon\alpha)} & 
      \mbox{ for}\quad R_{\rm min} > d/\varepsilon  .
    \end{array}
  \right.
\end{equation}   
The coefficients $c_T$ or $c'_T$ cannot
be determined within this scaling approach. This requires
taking into account contributions from the vdW interaction from all
distances, according to their statistical weight, instead of singling
out a the contributions from a typical distance $R_{\rm min}$.
We will quantify the coefficient $c_T$ within the functional RG
calculation presented in the next section.

\subsubsection{Functional Renormalization}
\label{sec_frg}

The problem that could not be handled by the approaches discussed
above is to properly account for the contributions from the vdW
interaction to the free energy on all length scales between $\lambda$
and $\ell$. 
This issue will be addressed now by a functional renormalization
group calculation. In particular, we will
use a procedure that is an extension of Wilson's approximate
recursion relation \cite{Wilson71}, and that has been well established
in the context of the wetting transition \cite{wetting}. 

We will only give a short sketch of the procedure here; a
comprehensive review can be found, e.g., in \cite{DombGreen}. The
starting point is the reduced Hamiltonian  $\cH\{\br\} \equiv
H\{\br\}/T$ defined by
\begin{equation}
\label{Hamiltonian}
  \cH\{\br\} =
  \int dz 
  \left[ 
    \frac{\varepsilon_\circ}{8T} 
    \left( 
      \frac{\partial \br(z)}{\partial z}
    \right)^2
    + \frac{V_0(\br(z))}{T} 
  \right]
\end{equation}
which describes a single line in $1+d'$ dimensions, interacting with a
stiff, straight line centered at $\br=0$ via the potential
$V_0(\br)$. $d'$ is the number of components of $\br$; in our case,
$d'=2$. An ultraviolet cutoff $\pi/\Lambda$, corresponding to the
shortest relevant length scale (parallel to the FL), is implied. This
problem is  
equivalent to the problem of two fluctuating FLs interacting with each
other, where $\br(z)\equiv {\bf s}_1(z)-{\bf s}_2(z)$ is their
difference coordinate. Accordingly, the line stiffness
$\varepsilon_\circ/4$ is equal to the single vortex line stiffness
$\varepsilon_l(k)\simeq \varepsilon_\circ/2$ 
divided by 2, where we have approximated the full, dispersive line
stiffness (\ref{stiff}) by its form  in the long wavelength limit
$\lambda k\ll 1$ as given by (\ref{stiffLongWave}), neglecting the
contribution $\sim\varepsilon^2\ll 1$.

In an RG step, the fast fluctuations $\br_>$ with wave vectors
between $\Lambda/b$ and $\Lambda$ are integrated out in an approximate
manner, yielding a renormalized interaction potential
$V'(\br_<)$.  $b>1$ is the rescaling factor, and $\br_<$ consists only
of Fourier modes with wave vectors in the range $0\ldots
\Lambda/b$. Then, one brings back the Hamiltonian to its original form
by rescaling $z\rightarrow z'=z/b$ and
$\br(z')\equiv\br_<(z=bz')/b^\zeta$, where $\zeta$ is the roughness
exponent, as defined in (\ref{displace}). For purely thermal 
fluctuations which we consider here, $\zeta=1/2$. For the interaction
potential, this implies the rescaling 
\begin{equation}
\label{V_rescaling}
  V^{(1)}[\br(z')] = b\,V'[b^\zeta \br(z')].
\end{equation}
This procedure is iterated until all fluctuations on scales smaller
than $\ell$ are integrated out. We denote the renormalized and
rescaled potential after the $N^{\rm th}$ iteration step with
$V^{(N)}(r)$. 

For $d'=2$ dimensions, the full nonlinear recursion relation is given by
\begin{equation}
\label{NonlinRecursion}
  V^{(N+1)}(\br) = -\tilde v \,b\, 
  \ln
    \int \frac{d^2r'}{2\pi\tilde a^2} 
    \exp\left(-\frac{\br'^2}{2\tilde a^2} - K(\br,\br')\right),
\end{equation}
where the kernel is defined as
\begin{equation}
  K(\br,\br') = 
  \frac{1}{2\tilde v}
  \left[ V^{(N)}(b^\zeta \br-\br') + V^{(N)}(b^\zeta \br+\br') \right].
\end{equation}
The length scale $\tilde a$, defined by $\tilde a^2 \equiv \langle
\br_>^2 \rangle$,  reads
\begin{equation} 
\label{def_tildea}
  \tilde a^2(b) = 
  \frac{8T}{\varepsilon_\circ} \int_{\Lambda/b}^\Lambda \frac{d\omega}{2\pi}\,
  \frac{1}{\omega^2} = \frac{4T\,(b-1)}{\pi\varepsilon_\circ\Lambda} 
  \equiv a^2(b-1),
\end{equation}
and the energy scale $\tilde v$ is given by
\begin{equation}
\label{def_tildev}
  \tilde v(b) = (1-b^{-1})\,v \qquad \mbox{with}\quad
  v = T\Lambda/\pi.
\end{equation}
$a$ and $v$ have been defined such that they do not depend on the
rescaling factor $b$, so it is convenient to choose them as natural
scales for length and energy, respectively.  The relation of these
scales $a$ and $v$ to the physically relevant scales $\lambda$ and
$\varepsilon_\circ=L_T T/\lambda^2$ takes the following form: For the
length scale we find $a^2 = (4/\pi^2) \,\lambda^2$.  Here, we have
identified the short scale cutoff $\pi/\Lambda$ from
(\ref{Hamiltonian}) with $L_T$ (cf.~Eq.\ \ref{L_T}), the segment
length on which the FL makes typical perpendicular excursions of order
$\lambda$, because we want to integrate out fluctuations of $\bu$ on
scales $\gtrsim\lambda$.  The resulting relation connecting $a$ and
$\lambda$ does {\em not} depend on $T$. In the following, we will
simply set $\lambda=a$. For the energy scale we find $v =
T^2/\varepsilon_\circ\lambda^2$ which {\em does} depend on
temperature.

Note that at low temperatures, $L_T$, the smallest length scale
parallel to the FLs in the bare Hamiltonian (\ref{Hamiltonian}) --
which we treat, by virtue of this choice of $\Lambda$, as an effective
Hamiltonian on the (perpendicular) scale $\lambda$ --, is
quite large: for $T=1\,$K it is as large as $2\,$cm. The applicability
of the RG is however restricted to samples whose height $L$ is much
larger than $L_T$ -- or, for a given sample, to temperatures $T$ such
that $L_T\ll L$. For smaller systems -- or smaller temperatures --
fluctuations of the FLs are smaller than $\lambda$, so it is a good
approximation to treat them as straight lines in the dilute limit
$\ell > \lambda$ we are interested in. With $L\ll L_T$, we are hence
in the disentangled flux liquid phase where practically no collisions
between FLs occur \cite{Nelson}. This means that also the contribution
from the entropic repulsion is absent, hence $\gamma_T=0$ in
(\ref{GibbsScaling}). Consequently, it is a good approximation in this
limit to replace the free energy density $F(\ell;T)$ in the Gibbs free
energy (\ref{GibbsGeneral}) by the bare FL interaction,
\begin{equation}
\label{G_LTggL}
  G(\ell;H,T) \simeq \frac{1}{\ell^2}
  \left[ \frac{z}{2}\, V_0(\ell) - \varepsilon_\circ\tilde h
    \ln\kappa\right],
  \qquad
  (L_T\gg L)
\end{equation}
where $V_0$ is defined as in (\ref{bareV0}). 
In the following, we assume that $L_T\ll L$. When discussing the
phase diagram, we will come back to the opposite case and use
expression (\ref{G_LTggL}) for the Gibbs free energy density in the
low temperature limit where $L_T\gg L$.

The renormalization procedure sketched above has been carried out
numerically. Restricting ourselves first to the extremely decoupled limit
$\varepsilon\rightarrow 0$, we define the bare interaction
(\ref{bareV0}) ${\cal V}^{(0)}(R)\equiv V_0(R)/v$ as
\begin{equation}
\label{bareV0spec}
  {\cal V}^{(0)}(R) = v_0 \left( K_0(R/\lambda) - a_{\rm vdw} f(R/\lambda)
    \frac{\lambda^4}{R^4} \right),
\end{equation}
where 
$v_0=2\varepsilon_0/v=2\,(\varepsilon_\circ\lambda/T)^2$, and the
strength of the thermal vdW attraction (\ref{vdwj}) is determined by
$a_{\rm vdw}=T/( 2d\varepsilon_\circ
\ln^{2}(\pi\lambda/d) )$. $f(x)$ is a function that smoothly 
cuts off the power law tail at $R\approx\lambda$, which we have
defined as
\begin{equation}
  f(x) = \left\{
    \begin{array}{ll}
      0, & x \le x_1 \\
      \frac14
        \left[ 
          1+\sin\left(\pi\frac{x-(x_1+x_2)/2}{x_2-x_1}\right)
        \right]^2, 
        & x_1<x<x_2 \\   
      1, & x \ge x_2
    \end{array}
  \right.
\label{CutOff}
\end{equation}
with $x_1=1$ and $x_2=5$. The choice of the cutoff function, as well as
the actual values of $x_1$ and $x_2$, is to some extent
arbitrary; $x_1=1$ is however an obvious choice, and $x_2$ has to be
chosen such that the cutoff is not too sharp and,
on the other hand, does not influence the form of the potential in
the vicinity of the minimum for those values of $a_{\rm vdw}$ that are
physically meaningful (cf.\ the discussion in section
\ref{sec_scaling}). We have carefully checked that the RG 
results are stable with respect to a variation of $x_2$.

The potential (\ref{bareV0spec}), with $v_0=1$, is shown
in Fig.~\ref{bare_pot}. For BiSCCO, one finds (from the discussion in
section \ref{sec_scaling}) a typical value $a_{\rm
  vdw}\approx 2\cdot 10^{-3}$ at $T=100$ K. With this van der Waals
amplitude, the bare potential $V_0(R)$ has a shallow minimum at
$R_{\rm min}\approx 18\lambda$; cf.\ the inset of Fig.~\ref{bare_pot}.

As a consistency check for the procedure, we have compared the
functional renormalization of a purely repulsive, short-ranged
potential under the action of the recursion relation
(\ref{NonlinRecursion}) with a known 
RG flow, namely the renormalization of a delta-like, repulsive
interaction of two elastic polymers by thermal fluctuations. In this
model, the amplitude $\gamma$ of the delta-like potential 
vanishes asymptotically like $\gamma^R(l)\sim 1/l$
\cite{NelsonSeung}. Here, $l$ is the continuous renormalization
parameter which is connected to our $N$ via $e^l = b^{\zeta N}$. To
make contact to this model, we have iterated the recursion relation
for the potential (\ref{bareV0spec}) with $a_{\rm vdw}=0$, so that the
bare potential is short ranged and purely repulsive, and
checked that the amplitude of the renormalized potential $V^{(N)}(R)$
does indeed scale like $V^{(N)}(0)\sim 1/N$ (the corresponding data
for $V^{(N)}(R)$ are not shown here; they look however similar to
those in the inset of Fig.~\ref{vNa3}, see below).

Now, we have iteratively applied the recursion relation
(\ref{NonlinRecursion}) to the potential (\ref{bareV0spec}) for different
values of the van der Waals amplitude $a_{\rm vdw}$ and global
amplitude $v_0$. One has some freedom in choosing the rescaling factor
$b$. Larger values for $b$ require less steps $N$ to reach the
same total rescaling factor $l=N\zeta\ln b$ and hence lead to less
numerical roundoff errors, at the price of a lower resolution in $l$.

\placefigure{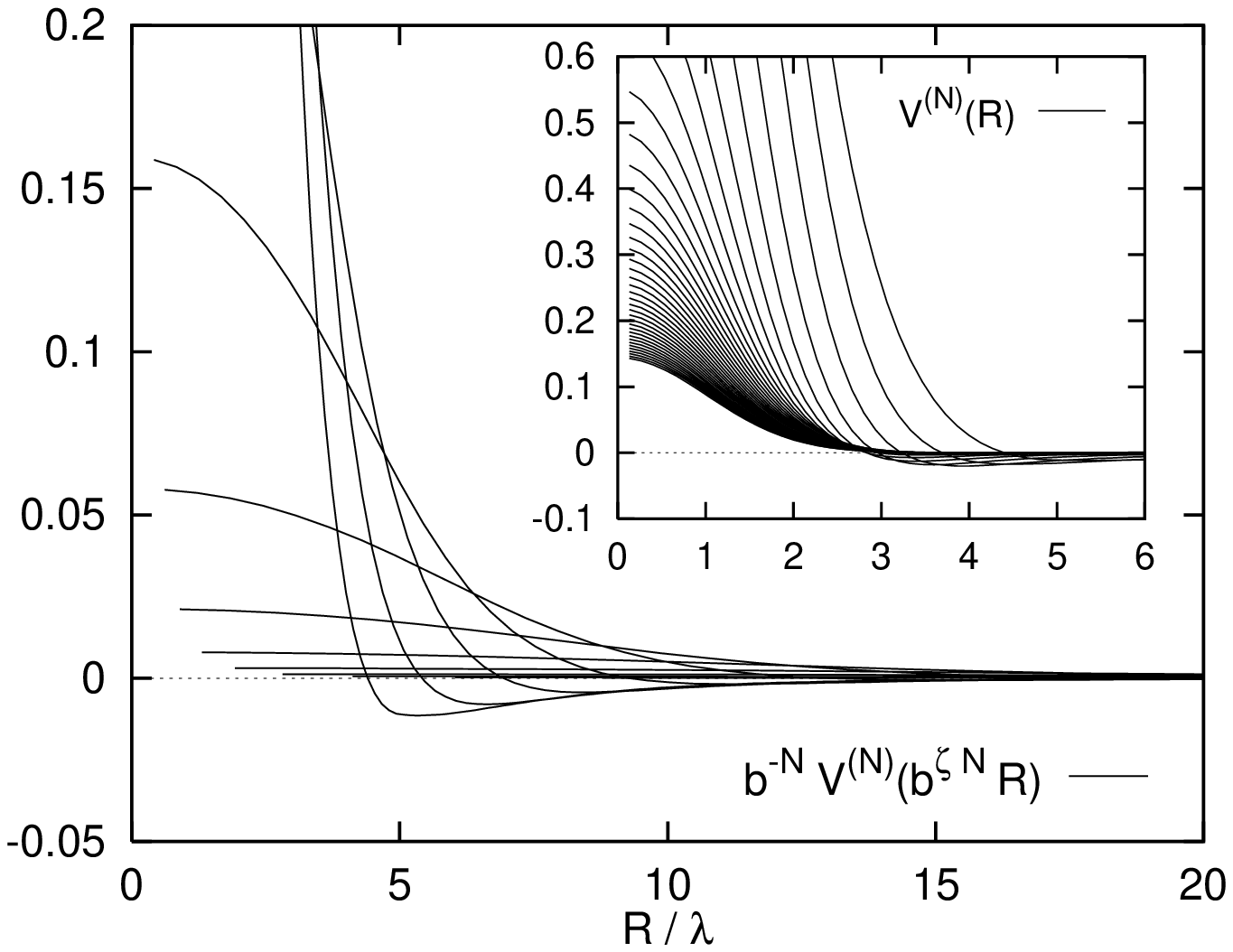}{0.48\textwidth}{ Functional renormalization of the
potential (\protect\ref{bareV0spec}) with $v_0=10$, van der Waals
amplitude $a_{\rm vdw}=3$ and a rescaling factor $b=1.1$. Only every
8$^{\rm th}$ step $N$ is shown. In the main figure, space and energy
have been scaled back to original space by plotting
$b^{-N}{\cal V}^{(N)}(b^{\zeta N}R)$. The bare potential with $N=0$
corresponds to the curve with the deepest minimum; for higher $N$, the
position of the minimum is shifted to larger values of $R$, and the minimum
becomes more shallow, while the amplitude in the origin decreases. In the
inset, ${\cal V}^{(N)}(R)$ as defined by (\protect\ref{NonlinRecursion})
is shown. Here, the rightmost curve is the bare potential.} 

\placefigure{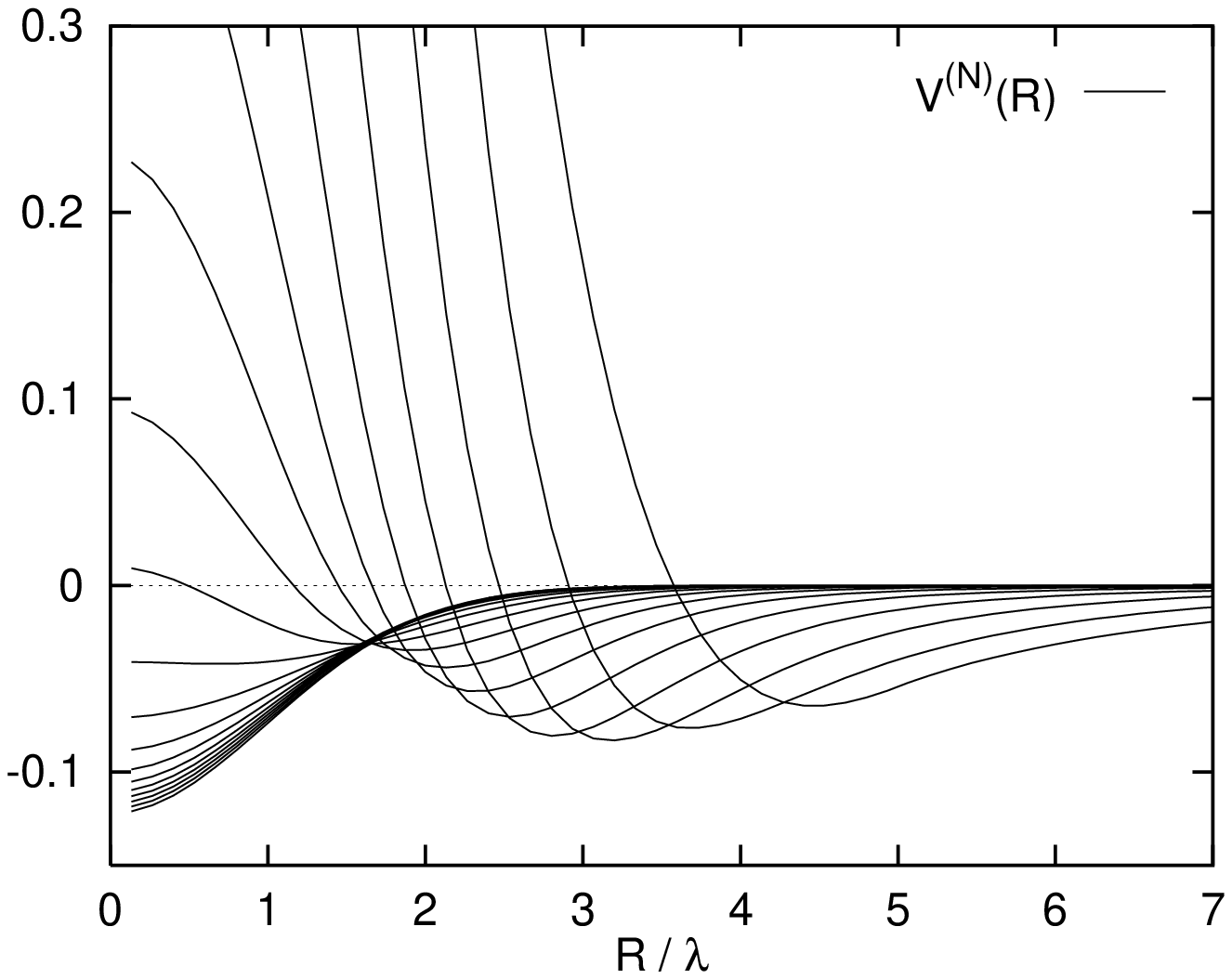}{0.4\textwidth}{The same plot as in the inset of
Fig.~\protect\ref{vNa3}, now with $a_{\rm vdw}=5.7$. The potential ${\cal
V}^{(N)}(R)$ is asymptotically mapped onto a purely attractive potential.
Again, the rightmost curve corresponds to the bare potential.}

We illustrate the procedure with data based on a bare potential
with fixed $v_0=10$, $b=1.1$ and varying $a_{\rm vdw}$.
The evolution of the potential $V^{(N)}(R)$ under the action of the
recursion relation (\ref{NonlinRecursion}) is illustrated
in Figs.~\ref{vNa3} and \ref{vNa5_7} for two different values for
$a_{\rm vdw}$. With these values, the bare
potential $V_0(R)$ belongs to two different generic cases: For
$a_{\rm vdw}<a^*_{\rm vdw}$, where $a^*_{\rm vdw}$ is a critical 
value that depends on $v_0$ and, very weakly, on $b$, ${\cal V}^{(N)}$
is asymptotically mapped onto a purely repulsive
short range potential, see Fig.~\ref{vNa3}. For
$a_{\rm vdw}>a^*_{\rm vdw}$, on the other hand, ${\cal V}^{(N)}$ is
mapped onto a purely attractive potential, as shown in
Fig.~\ref{vNa5_7}.

\subsubsection{Gibbs free energy density}

Now, let us come back to our starting point: The calculation of the
effective Gibbs free energy density (\ref{GibbsGeneral}) which we rewrite as 
\begin{equation}
\label{GibbsMu}
  G(\ell;\mu,T) = (V^{\rm eff}(\ell) + \mu) / \ell^2,
\end{equation}
where
$\mu=-\varepsilon_\circ\tilde h\ln\kappa$
plays the role of a chemical potential.
$G(\ell;\mu,T)$ has to be minimized as a function of $\ell$ for fixed
$\mu$.  The effective potential is given by 
\begin{equation}
\label{Veff_def}
  V^{\rm eff}(\ell) \equiv b^{-N} V^{(N)}(\lambda),
\end{equation}
where $N$ is determined from $\ell=\lambda b^{\zeta N}$. With this
$N$, the renormalized $V^{(N)}$ includes thermal fluctuations on scales 
between $\lambda$ and the mean flux line distance $\ell$. It is evaluated
at the (rescaled) position $\lambda$, corresponding to $\ell$ in
unrescaled coordinates.

The effective potential is plotted in Fig.~\ref{Veff} for some values
of $a_{\rm vdw}$. For $a_{\rm vdw}>a^*_{\rm vdw}$, 
where $a^*_{\rm vdw}\approx 5.33$ with the other parameters $v_0$ and
$b$ set to the
values given above, $V^{\rm eff}(\ell)$
has a global minimum at some finite value $\ell_{\rm min}$. For
$\ell\gg \ell_{\rm min}$, it decays $\sim -1/\ell^2$. This confirms
the prediction from (\ref{ScalingVDW}) that the contribution
from the vdW interaction to the free energy {\em density} scales like
$-1/\ell^4$ in the thermal case. 

\placefigure{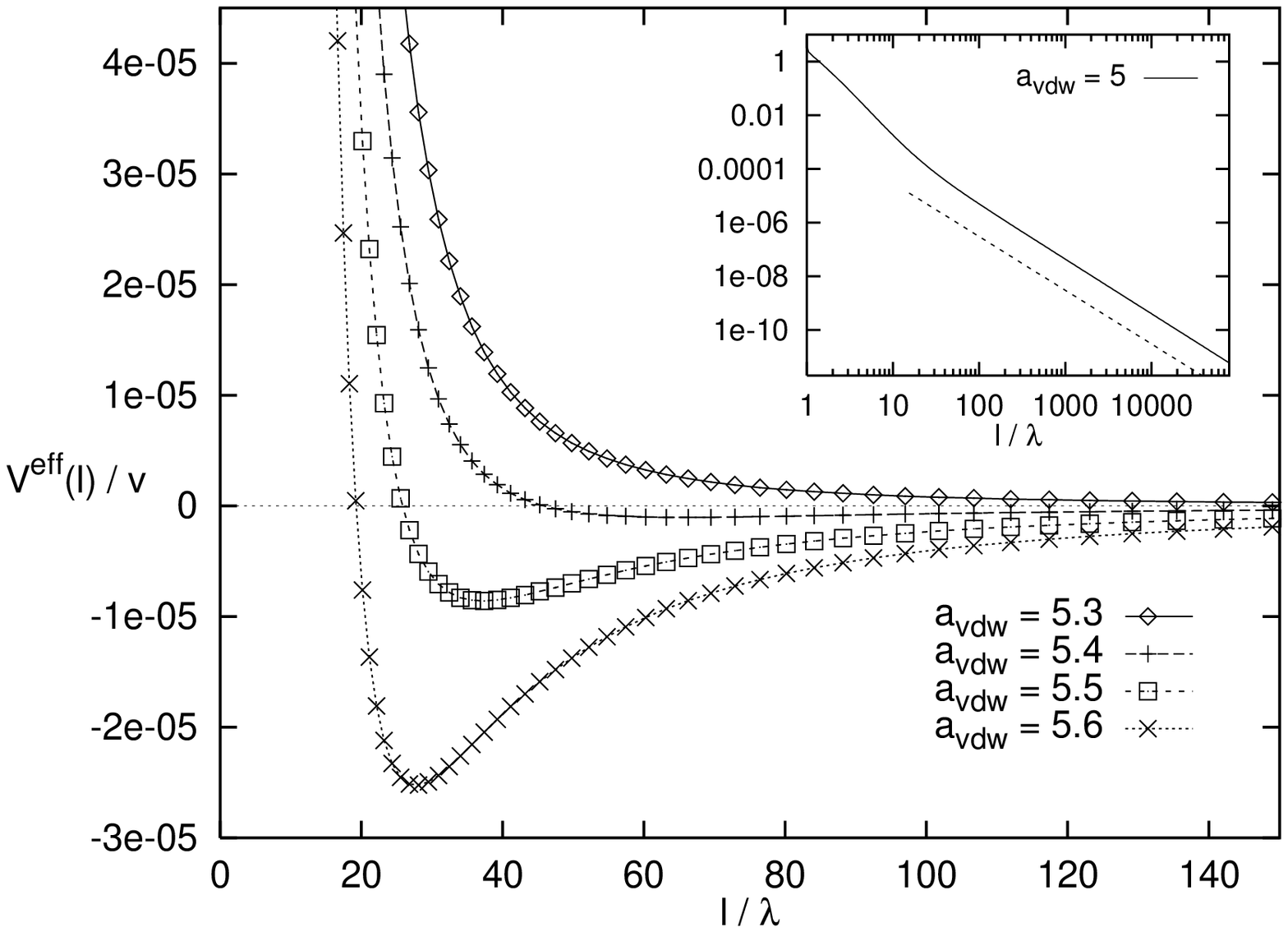}{0.48\textwidth}{Effective potential potential
  $V^{\rm eff}(\ell)$ (\protect\ref{Veff_def}) for rescaling factor
  $b=1.1$, potential amplitude $v_0=10$ and several values for the
  van der Waals amplitude $a_{\rm vdw}$ in the vicinity of $a^*_{\rm
  vdw}$. The  inset shows a double logarithmic plot of a purely
  repulsive $V^{\rm eff}(\ell)$ with $a_{\rm vdw}=5.0<a^*_{\rm
  vdw}$. For comparison, a (dashed) line proportional to $1/\ell^2$
  has been   added. } 

For $a_{\rm vdw}<a^*_{\rm vdw}$, on the other hand,
$V^{\rm eff}(\ell)$ is purely repulsive and decays 
$\sim 1/\ell^2$ for
large $\ell$ (cf.\ the inset of Fig.~\ref{Veff}), revealing that the
entropic contribution $\sim T/\ell^2$ to the Gibbs free energy dominates over
the vdW attraction.

$G(\ell;\mu,T)$ is now minimized
as a function of $\ell$ for fixed chemical potential $\mu$. The 
corresponding data for $G$ as a function of $\ell$ are shown in Fig.\
\ref{Gibbs} for two typical values of $a_{\rm vdw}$ smaller and 
larger than $a^*_{\rm vdw}$, respectively. For $a_{\rm vdw}<a^*_{\rm
  vdw}$, the minimum position $\ell_0$ is continuously 
shifted to larger values when $\mu\rightarrow 0$ from below. This
corresponds to a second order phase transition at $\mu=0$, with
$B=\Phi_0/\ell^2_0$ 
vanishing continuously for $H\searrow H_{c_1}^0$. Analysis of
$\ell_0$ as a function of $\mu$ reveals $B\sim 
\mu\,\ln(1/\mu)$, hence reproducing the result from the
renormalization group treatment in \cite{NelsonSeung}.

For $a_{\rm vdw}>a^*_{\rm vdw}$, the position of
the minimum remains at a finite position determined by the form of
$V^{\rm eff}(\ell)$ when $\mu$ approaches 0 from below. This minimum
disappears at $\mu=\mu^*>0$; for $0<\mu<\mu^*$, $G$ has two
minima, one at a finite length $\ell$ and the other one at
infinity. This scenario describes a first order phase transition with
$B$ dropping to zero from a finite value $B_v$ when $H\searrow
H_{c_1}^r \equiv H_{c_1}^0-4\pi\mu^*/\Phi_0$.

\placefigure{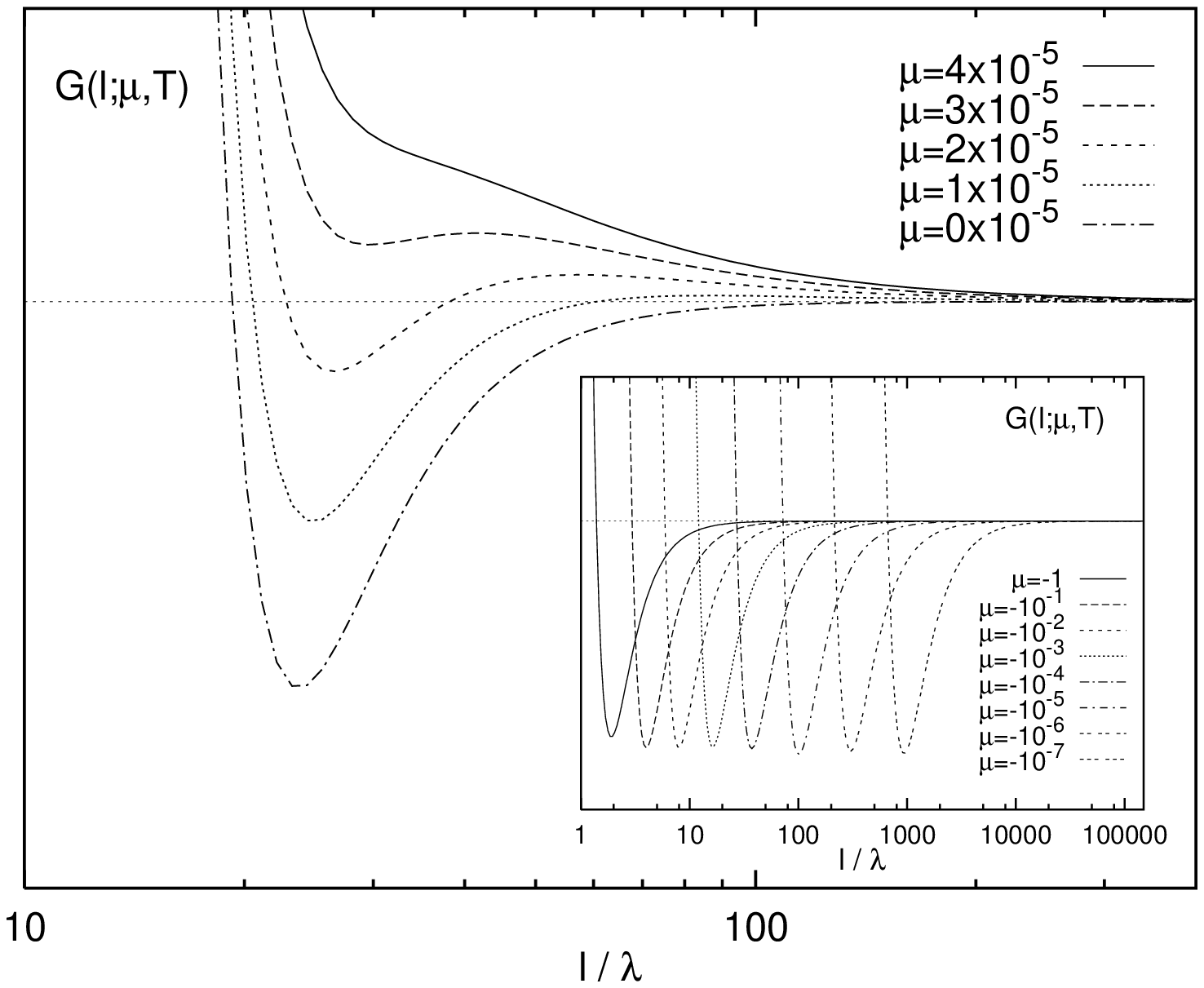}{0.48\textwidth}{{\em Main figure:} Gibbs free
  energy density $G(\ell;\mu,T)$ 
  (\protect\ref{GibbsMu}) as a function of the mean flux line spacing
  $\ell$ for different values $\mu$, based on functional RG data
  starting from a 
  bare potential (\protect\ref{bareV0spec}) with $a_{\rm vdw}=5.6>a^*_{\rm
  vdw}$, $v_0=10$ and $b=1.1$. The effective potential $V^{\rm
  eff}(\ell)$ for this vdW 
  amplitude is shown in Fig.~\protect\ref{Veff}. Only data from
  the regime $0<\mu\protect\lesssim\mu^*$ where $G$ has two minima are
  shown. The top curve corresponds to the largest value of $\mu$.
  {\em Inset:} The same plot, now for data from a bare
  potential with $a_{\rm vdw}=5.0<a^*_{\rm vdw}$. The different curves
  have been individually scaled for the plot. From left to
  right, the absolute value of $\mu$ becomes smaller.}

Now, we compare the results from the functional RG calculation with
the  effective Gibbs free energy density derived by scaling arguments in
section \ref{sec_scaling}. As stated above,
the scaling behavior for large $\ell$ given by
expression (\ref{GibbsScaling}) is confirmed by the RG results both in the
case where the entropic repulsion dominates ($\delta_T<\gamma_T$, 
corresponding to $a_{\rm vdw}<a^*_{\rm vdw}$) and in the opposite case
where the vdW attraction gives the larger contribution.
At the temperature $T^*$ where
$\gamma_{T^*}=\delta_{T^*}$, one has $a_{\rm vdw}=a^*_{\rm vdw}$. This
notion allows for a straightforward calculation of the coefficient
$c_T$, which determines the amplitude of the vdW contribution to the
free energy. Using the definitions for $v_0$ and $a_{\rm vdw}$ as given after
Eq.~(\ref{bareV0spec}), together with (\ref{gamma_T}) and (\ref{delta_T}) in
the limit $R_{\rm min} < d/\varepsilon$, this leads to
\begin{equation}
\label{c_T}
  c_T \approx 9\, \frac{\alpha^4}{\beta^2} 
  \,\frac{1}{v_0\,a^*_{\rm vdw}}.
\end{equation}

We have determined $a^*_{\rm vdw}$ numerically for a large range of
values for $v_0=10^4\ldots 10^7$. Here and below, we will report
results from functional RG calculations using a rescaling factor $b=1.5$. 
It turns out that the right hand side of (\ref{c_T}) is indeed
basically constant if one defines the size of the minimum width to be 
half the minimum position, $\beta\equiv\alpha/2$ (which is quite an
obvious choice after inspection of the form of the bare potential for
different vdW amplitudes), and uses
$\alpha=R_{\rm min}/\lambda$ with $R_{\rm min}$ denoting the true
minimum position for a given value of $a^*_{\rm vdw}$ (for the given
range of $v_0$, $\alpha$ varies between 12 and 19). In this way, we
obtain $c_T=3.9\pm 0.3$. The data for the critical amplitudes of
the vdW interaction $a^*_{\rm vdw}$, as well as
the values for $c_T$ calculated from these data, are shown
in Fig.~\ref{V0vsAvdw} as a function of $v_0$. 

Accordingly, we have determined the prefactor $c'_T$ in the case of
finite anisotropy $\varepsilon>0$ where $R_{\rm min}
>d/\varepsilon$, corresponding to the lower term in
(\ref{delta_T}). We find a value of $c'_T=1.9\pm0.2$.

\placefigure{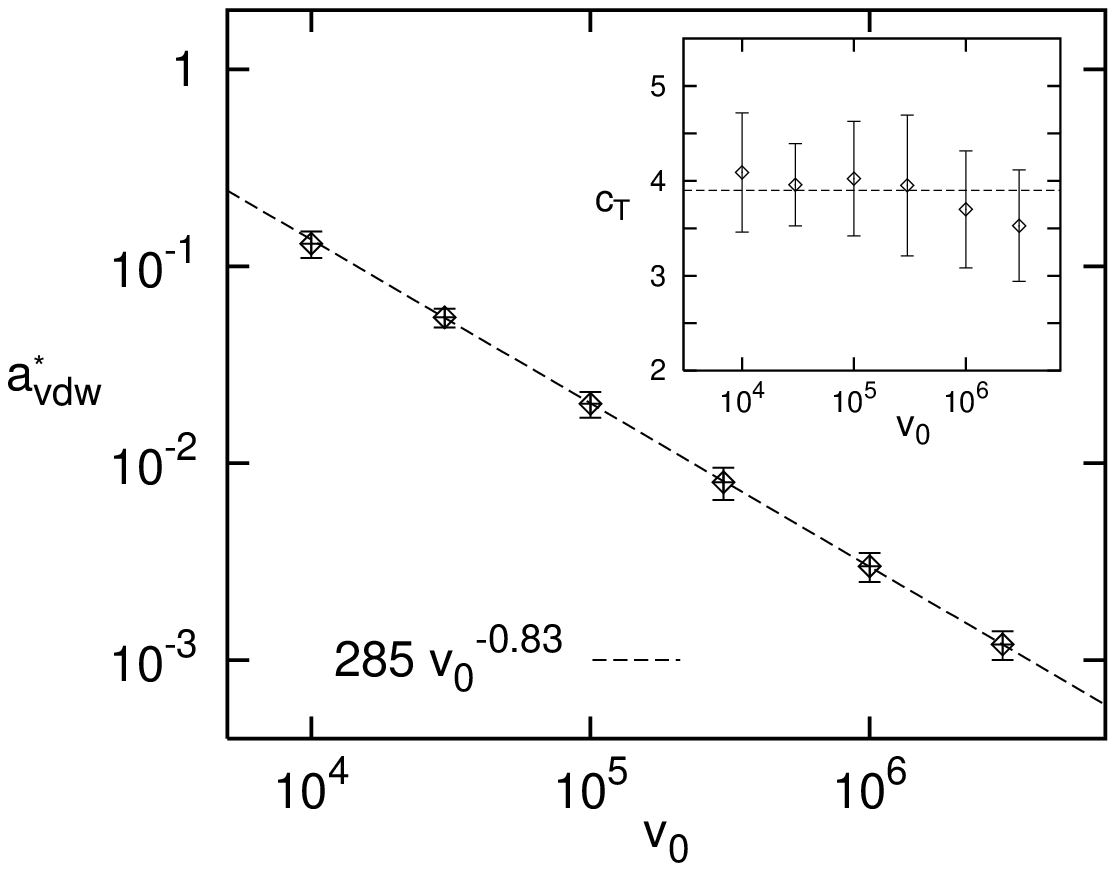}{0.4\textwidth}{'Critical' van der Waals
  amplitude $a^*_{\rm vdw}$ as a function of $v_0=2(\varepsilon_\circ
  \lambda/T)^2$. The product $v_0 a^*_{\rm vdw}$ is not a constant;
  instead, $a^*_{\rm vdw}\sim v_0^{-0.83}$ in the range shown in this
  figure. The values for $c_T$ calculated from these data are shown in
  the inset. The straight line corresponds to $c_T=3.9$.}

Now, we will discuss the consequences for the low field phase diagram
of layered superconductors. We restrict ourselves here to an analysis
based on the form
(\ref{vdwj}) of the thermal vdW interaction which decays with the
fourth power of $R$ and, consequently, to the upper definition of
$\delta_T$ in (\ref{delta_T}). We have also calculated the phase diagram
accordingly using expression (\ref{vdwk}) for the vdW energy
which is valid for $R>d/\varepsilon$. Since the qualitative features
are the same and even the numerical values are very close to each
other in the two different cases, we only demonstrate the data for the
first case. 

We have used the parameters for BiSCCO as given above
(which implies $a_{\rm vdw}\approx 2\cdot 10^{-5} \,T$/K), and have
determined the position $\ell_0$ of 
the minimum of the Gibbs free energy $G(\ell;\mu=0,T)$ which has been
calculated from the functional RG as described above. In
Fig.~\ref{phasediag}, the resulting data for the magnetic induction
$B_v=\Phi_0/\ell_0^2\approx 500\,$G$/(\ell_0/\lambda)^2$
\cite{blatrev} are shown; the corresponding line is labeled
$B_v^{_{\rm (RG)}}$. 

For comparison, we have determined the same field $B_v$ by minimizing
the Gibbs free energy (\ref{GibbsScaling}) as derived from 
the scaling argument, using (\ref{gamma_T}) and (\ref{delta_T}) for the
parameters $\gamma_T$ and $\delta_T$, with
$c_T=3.9$.  These data are shown in the same figure, denoted as
$B_v^{\rm (sc)}$. While the resulting values are of the
same order of magnitude as those obtained from the RG, $B_v^{_{\rm
    (RG)}}$, there is considerable quantitative disagreement. Hence,
although 
the scaling approach to the Gibbs free energy density $G(\ell;\mu,T)$
leads to the correct asymptotic scaling of the various contributions,
and even 
yields estimates for $B_v$ in the correct order of magnitude, it seems
to fail for the exact determination of the minimum of the Gibbs free
energy. Evidently, also the functional RG approach
is not exact: The derivation of
(\ref{NonlinRecursion}) contains some severe approximations, and
different choices of the rescaling factor $b$ lead to slightly
different numerical results. Nevertheless, the RG represents the most
natural and adequate method for the calculation of the effective Gibbs
free energy. 

\placefigure{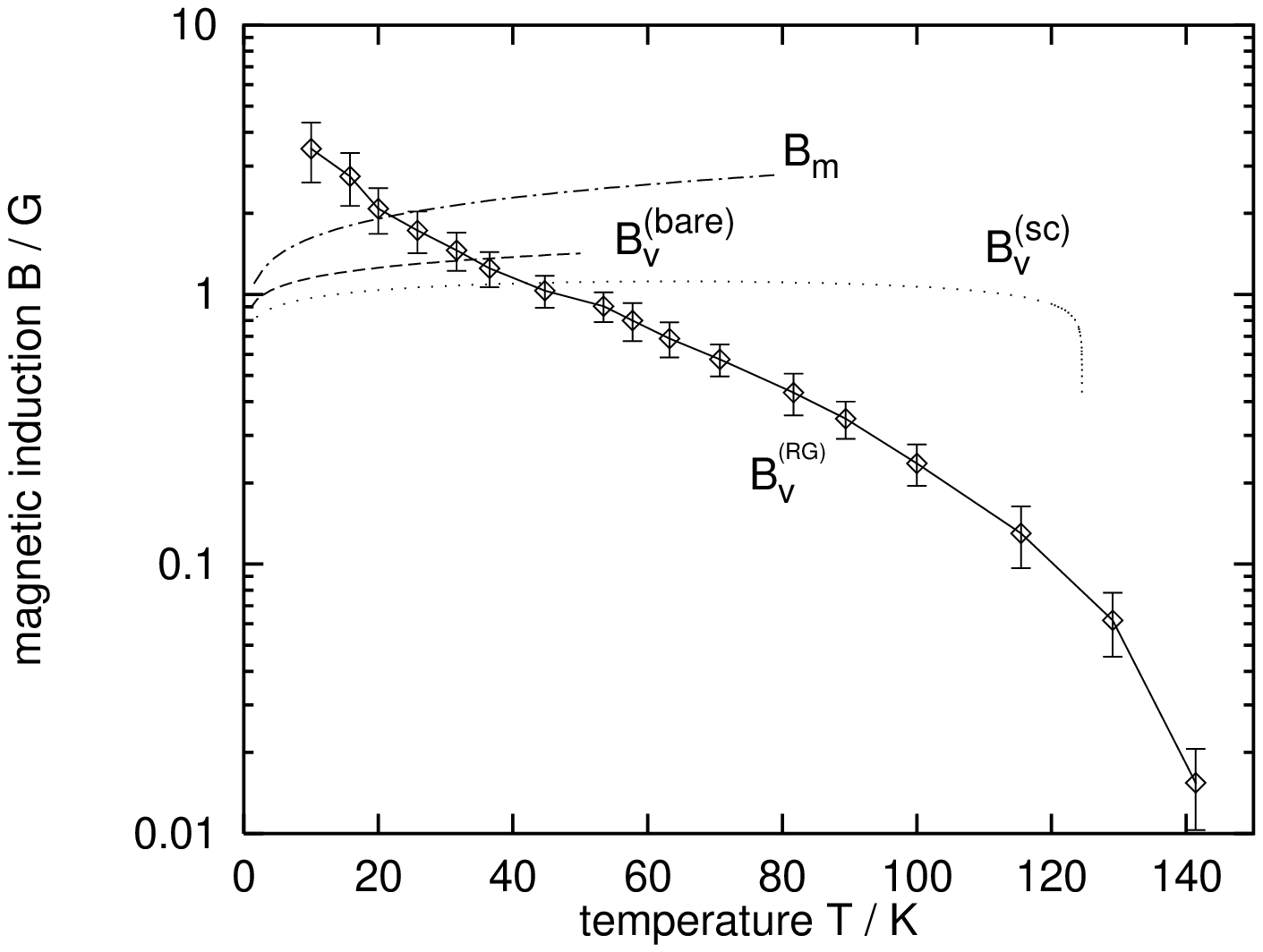}{0.45\textwidth}{Low field phase diagram for a
  layered superconductor; physical parameters typical for BiSCCO have
  been used. At low temperatures the Meissner-Ochsenfeld phase becomes
  unstable towards a flux liquid by a first order transition where $B$
  jumps to $B_v^{_{\rm (RG)}}$, which has been determined from the
  functional RG data for the effective interaction energy. For
  comparison, the same line $B_v^{\rm (sc)}$ as determined from the
  Gibbs free energy (\protect\ref{GibbsScaling}) that was derived from
  scaling arguments are shown as a dotted line. For very low
  temperatures, instead of $B_v^{_{\rm (RG)}}$ the ({\em dashed}) line
  $B_v^{(bare)}$ will correctly describe the jump in the magnetic
  induction. Finally, $B_m$ ({\em dashed-dotted line}) denotes a first
  order melting transition between the liquid and a solid phase at
  higher magnetic fields $B$.}

As discussed above (in section \ref{sec_frg}), the RG approach makes
sense only for temperatures such that $L_T$ is smaller than the sample
height $L$. For lower temperatures, corresponding to the
disentangled flux liquid phase, the bare interaction enters the 
Gibbs free energy, cf.\ Eq.~(\ref{G_LTggL}). The jump in the
magnetic induction that results from this expression is shown in
Fig.~\ref{phasediag}, labeled $B_v^{\rm (bare)}$. In contrast to the RG
result $B_v^{_{\rm (RG)}}$ (which grows for $T\rightarrow 0$), 
$B_v^{\rm (bare)}$ drops to 0 in this limit. The exact temperature beyond
which the RG result applies will depend on the sample height $L$, and
in a large temperature range around this value, the real line $B_v$
will lie somewhere between $B_v^{_{\rm (RG)}}$ and $B_v^{\rm (bare)}$.

The last line we have included in the phase diagram is the lower
(first order) melting line $B_m(T)$. This line separates the flux liquid
phase for $B<B_m$ from a solid phase for $B>B_m$, where the FLs are
organized in an Abrikosov lattice. For layered
superconductors, where the electromagnetic interactions dominates over
the Josephson coupling, this line is given by 
\begin{equation}
\label{B_m}
  B_m(T) = 
  \frac{\Phi_0}{4\lambda^2} \,
  \ln^{-2}
  \left[ 
    \frac{4\pi \, c_L^2}{(3\pi)^{1/4}} \,
    \frac{\varepsilon_\circ\lambda}{T} 
  \right]
\end{equation}
where $c_L\approx 0.3$ is the Lindemann number \cite{BlatterMelting}.
Note that $B_m>B_v^{_{\rm (RG)}}$ for a large temperature range; only
for very low temperatures, the two lines cross. For samples with finite
height $L$, however, for these low temperatures $B_v$ will cross over
to the line $B_v^{\rm (bare)} < B_m$, so that it is likely that $B_v<
B_m$ in the whole temperature range.

Finally, we want to compare our phase diagram with that proposed by 
Blatter and Geshkenbein \cite{BlatterGeshkenbein,bg2}.
While both proposals basically agree, including the
finding that $B_v$ is of order $1\,$G for low temperatures, there are some
differences. In particular, these authors find a bubble-like shape
of the instability region which is bounded by two lines $B_v$ and
$B_e$, where both $B_v$ and $B_e$ are finite for some temperature
range. In terms of the Gibbs free energy, this corresponds to the
simultaneous existence of two minima at finite values $\ell$, which we
don't find in the effective potential as derived from the functional RG
procedure, as can be seen for example in Fig.~\ref{Gibbs}.

\subsection{Phase diagram in the presence of disorder}
\label{subsec:disorder}

While the functional RG  presented in the preceding sections
only works for the thermal case, 
the scaling approach to the Gibbs free energy density
$G(\ell;H,T,T_{\rm dis})$ from section
\ref{sec_scaling} can easily be extended to the case with disorder.
The contribution from the disorder induced vdW attraction can be
estimated in the same way as that from the thermally induced vdW attraction,
which leads to Eq.~(\ref{ScalingVDW}),
where the value $\zeta=5/8$ for the roughness exponent has to be used instead
of the thermal value 1/2.
In contrast to the thermal case, where the
entropic repulsion and the effective vdW attraction have the same dependence
on the mean distance $\ell$, the resulting contribution from the disorder
induced vdW attraction decays faster than the (disorder induced) steric
repulsion.

The resulting expression for the Gibbs free energy density reads
\cite{NattermannMukherji}
\begin{equation}
  G(x;H,T,T_{\rm dis})\approx G(x;H,T,0) +
  \frac{\varepsilon_\circ}{\lambda^2x^2}
  \left\{
      \frac{\gamma_{\rm dis}}{x^{6/5}}
    - \frac{\delta_{\rm dis}}{x^{8/5}}
  \right\}
  \, .
\end{equation}
The amplitudes of the contribution from the steric repulsion and from the vdW
attraction are given by 
\begin{equation}
  \gamma_{\rm dis} \approx c_{\rm dis} \,\kappa^{4/5}
  \left( T_{\rm dis}/\varepsilon_\circ\lambda \right)^2
\end{equation}
and, for $R_{\rm min} \gg d/\varepsilon$,
\begin{equation}
 \delta_{\rm dis} \approx  
  {\tilde c}_{\rm dis}
  \frac{\beta^{8/5}}{\varepsilon\kappa^2\alpha^{5}}
  \left(
    \frac{T_{\rm dis}\kappa^2}{\varepsilon_\circ\lambda}
  \right)^{3\eta}
  \frac{1}{\ln^{1+2\eta} (\frac{\pi}{\alpha\varepsilon}) }
  \left(
    \frac{\lambda}{d}
  \right)^{\eta-1}  ,
\end{equation}
respectively, where $\alpha$ and $\beta$ are again the minimum
position and width of the bare potential, divided by $\lambda$, and
the exponent $\eta$ was defined in (\ref{eta}).

We briefly recall the essential features of the resulting phase  
diagram which has been considered semi-quan\-ti\-ta\-tively in
Ref.~\cite{NattermannMukherji}. 
The phase diagram in the presence of impurities is expected to  
differ significantly in the low temperature region $T<T_{\rm dis}$
where the   
disorder mediated fluctuations dominate the thermal fluctuations. At  
$T=0$,  as the magnetic field is increased for weak disorder (but  
$\Delta >1$) there is a continuous transition from the Meissner  
phase to the low density phase followed by a jump to a high density  
phase of FLs as the magnetic field is further increased. This  
behavior smoothly crosses over to the known thermal phase diagram.  
As the strength of the disorder increases the disorder induced  
repulsive interaction dominates and the first order transition at  
$T=0$ slowly disappears but still persists at finite low  
temperatures.  For even stronger disorder the first order transition  
completely vanishes.

In order to determine the dimensionless coefficients $c_{\rm dis}$ and
${\tilde c}_{\rm dis}$, in principal a renormalization procedure similar to
that applied in the thermal case can be used. Due to the presence of
disorder and, hence, the existence of metastable states,
however, this task is considerably more complicated. In a closely
related model, namely the disorder-induced unbinding of a flux line
from an extended defect, a functional RG calculation has
been applied in \cite{BalentsKardar}; it would
be interesting to see whether the method that was developed there can be
successfully applied in the present case.

\begin{acknowledgement}    
T.N.\ acknowledges the support of the Volkswagen-Stiftung as well as the 
hospitality of the Laboratoire de Physique th\'{e}orique of the ENS, Paris,
where some of the work was done. S.M.\ acknowledges support of  Deutsche
Forschungsgemeinschaft (SFB 341), and  A.V.\ acknowledges support of the
German-Israeli Foundation (GIF). 
\end{acknowledgement}

\begin{appendix}

\section{Probability density $p_{12}(\delta\varepsilon)$}
\label{AppLevelStat}

Consider two independent sets $S_i$, each consisting of $N$ Gaussian
random numbers $\varepsilon_m^{_{(i)}}$, with $i$=1,2. The random numbers
obey the Gaussian distribution
\begin{equation}
\label{Gaussian}
  f(\varepsilon) = \frac{1}{\sqrt{2\pi}\,\bar\varepsilon}\;
  e^{\ds -\varepsilon^2/2\bar\varepsilon^2}. 
\end{equation}
Furthermore, define 
$F(\varepsilon) = \int_{-\infty}^{\varepsilon} d\varepsilon'\,
f(\varepsilon')$ 
which, in this case of a Gaussian distribution, is an error function. 
For a given realization, let $\varepsilon_N^{_{(i)}}$ be the maximal
number in $S_i$. (Note that by symmetry, the following derivation also
applies to the minima.) 
The probability distribution for $\varepsilon_N$ is given by
\cite{NattermannRenz} 
\begin{equation}
  g_{N}(\varepsilon_N) = N F^{N-1}(\varepsilon_N)f(\varepsilon_N).
\end{equation}
Let $p_{12}(\delta \varepsilon)$ be the probability that the
difference between the maximal values in $S_1$ and $S_2$ is equal to
$\varepsilon_N^{_{(1)}}-\varepsilon_N^{_{(2)}}=\delta \varepsilon$. It is
given by the convolution 
\begin{equation}
\label{p12_def}
  p_{12}(\delta \varepsilon) = \int_{-\infty}^{\infty}
  d\varepsilon\;g_{N}(\varepsilon)\,g_{N}(\varepsilon+\delta \varepsilon).
\end{equation}
This integral can be evaluated by a saddle point approximation,
which results in
\begin{equation}
\label{p12_saddlepoint}
  p_{12}(\delta \varepsilon) \approx
  \frac{\ln^{1/2} N}{\sqrt{2\pi}\; \bar\varepsilon} \, 
  \exp\left(
    -\frac{\ln N}{2} \frac{ \delta \varepsilon^2 }{ \bar\varepsilon^2 }
  \right).
\end{equation}
This Gaussian approximation is good for values in the range $|\delta
\varepsilon|\lesssim \bar\varepsilon$, which is the regime we are
mainly interested in. For  $\delta \varepsilon\gg
\bar\varepsilon$, on the other hand, $p_{12}(\delta
\varepsilon)\simeq N f( \overline{\varepsilon_N}  +\delta
\varepsilon),$ where $\overline{\varepsilon_N} =\int d\varepsilon
\,\varepsilon \,g_{N}(\varepsilon)$. For the Gaussian
(\ref{Gaussian}), $\overline{\varepsilon_N}\simeq
  c\,\bar\varepsilon\,\ln^{1/2}N$ with $c\approx 2\sqrt\pi/e$.

\vspace*{6cm}

\section{List of symbols}

\begin{tabular}{llr}
\hline\hline 
\rule{0cm}{3ex}
$a_{\rm vdw}$ 
          & reduced vdW amplitude & 
            (\ref{bareV0spec}) \\
$B$       & magnetic induction $B=\Phi_0/\ell^2$ \\
$B_m$     & lattice melting field & (\ref{B_m}) \\
$B_v$     & van der Waals transition field  & \\
$c_T{}^{(}{\!'}{}^{)}$     & dimensionless factor in the vdW & \\
          &  contribution to the free energy & (\ref{delta_T}) \\          
$C_T$     & connected correlation function & (\ref{C_T}) \\
$C_{\rm dis}$ & disconnected correlation function &
   (\ref{C_dis}) \\ 
$d$       & layer distance & \\
$\delta_T$ & amplitude of the vdW contribution \\
          &  to the free energy &           (\ref{GibbsScaling}) \\
$\Delta_0$ & reduced disorder strength & (\ref{delta}) \\
$\Delta(x)$  & wave vector dependent reduced dis-  & \\
          & order strength  &    (\ref{delta}) \\ 
$\Delta$  & reduced disorder strength for pan- \\
          & cake vortices $\Delta=\Delta(\pi\lambda/d)$ & \\
$\varepsilon$ & anisotropy parameter & \\
$\varepsilon_\circ$ & basic energy scale
            $\varepsilon_\circ=(\Phi_0/4\pi\lambda)^2$ \\ 
$\varepsilon_l$ & line stiffness of a single FL & (\ref{stiff})
\\
$\varepsilon_{\rm pin}(\br,z)$ & random pinning potential &
   (\ref{mutual}) \\ 
$\varepsilon_{\rm pin}(\bu)$ & pancake vortex pinning potential & 
   (\ref{epspin_d}) \\
$\bar\varepsilon$ 
          & disorder scale & (\ref{eps_variance}) \\
$\gamma_T$& amplitude of entropic repulsion & (\ref{gamma_T}) \\
$G$       & Gibbs free energy density & (\ref{GibbsGeneral}) \\
$H$       & external magnetic field & \\
$H_{c_1}$ & lower critical field & \\
$\kappa$  & GL parameter $\kappa=\lambda/\xi$ & \\
$\ell$    & mean distance between FLs& \\
$\lambda$ & London penetration depth & \\
$L$       & total flux line length & \\
$L_T$     & thermal length scale $L_T=\varepsilon_\circ \lambda^2/T$ 
          & (\ref{L_T}) \\
$L_{\rm dis}$ & disorder length scale & (\ref{displace}) \\
$\mu$     & 'chemical potential' $\mu=-\varepsilon_\circ \bar h
          \ln\kappa$ & (\ref{GibbsMu}) \\
$\Phi_0$  & unit flux quantum $\Phi_0=hc/2e$& \\
$T_{\rm dis}$ & disorder temperature & \\
$\bu$     & vortex displacement & (\ref{mutual}) \\
$u^2_{\rm pv}$ & typical pancake vortex fluctuations & (\ref{correl1}) \\
$U_{12}$  & dipole-dipole interaction& (\ref{U_12}) \\
$v$       & basic energy scale in the RG & (\ref{def_tildev}) \\
$v_0$     & $=2\varepsilon_\circ/v$, reduced interaction energy
          & (\ref{bareV0spec}) \\
$V_{\alpha\beta}^{\rm int}$ 
          & \multicolumn{2}{l}{London potential \hfill
          (\ref{iso})-(\ref{V_xx_eps0})} \\ 
$V_0$     & total bare FL interaction  & (\ref{bareV0}) \\
$V_{\rm vdw}^{\rm th}$ & 
            \multicolumn{2}{l}{thermal vdW interaction \hfill 
              (\ref{vdwj}), (\ref{vdwk})} \\
$V_{\rm vdw}^{\rm dis}$ 
          & disorder vdW interaction 
          & (\ref{vdwdis1}) \\        
$\xi$     & coherence length & \\
$\langle\ldots\rangle$ & thermal average & \\
$\;\overline{\ldots\rule{0cm}{1ex}}$    & disorder average & \\ \hline\hline
\end{tabular}

\end{appendix}

\end{document}